\input harvmac.tex
\input amssym.tex

%\draftmode

\font\teneurm=eurm10 \font\seveneurm=eurm7 \font\fiveeurm=eurm5

\newfam\eurmfam

\textfont\eurmfam=\teneurm \scriptfont\eurmfam=\seveneurm

\scriptscriptfont\eurmfam=\fiveeurm

 \font\teneusm=eusm10 \font\seveneusm=eusm7 \font\fiveeusm=eusm5

\newfam\eusmfam

\textfont\eusmfam=\teneusm \scriptfont\eusmfam=\seveneusm

\scriptscriptfont\eusmfam=\fiveeusm

\font\tencmmib=cmmib10 \skewchar\tencmmib='177

\font\sevencmmib=cmmib7 \skewchar\sevencmmib='177

\font\fivecmmib=cmmib5 \skewchar\fivecmmib='177

\newfam\cmmibfam

\textfont\cmmibfam=\tencmmib \scriptfont\cmmibfam=\sevencmmib

\scriptscriptfont\cmmibfam=\fivecmmib

%$\varpi,\vartheta,\varrho,\varsigma,\Xi$

%$$\eusm{ABCDEFGHIJKLMNOPQRSTUVWXYZ}$$

%%%%%%%%%%%%%%%%%%%%%%%%%%%  REFERENCES  %%%%%%%%%%%%%%%%%%%%%%%%%%%%%

\lref\Nekrasov{N.~Nekrasov,
``Seiberg-Witten Prepotential From Instanton Counting,''
Adv.\ Theor.\ Math.\ Phys.\  {\bf 7} (2004) 831-864, hep-th/0206161.}

\lref\IqbalV{T.~J.~Hollowood, A.~Iqbal, C.~Vafa,
``Matrix models, geometric engineering and elliptic genera,'' hep-th/0310272.}

\lref\GSV{S.~Gukov, A.~Schwarz, C.~Vafa,
``Khovanov-Rozansky homology and topological strings,''
Lett.\ Math.\ Phys.\ {\bf 74} (2005) 53, hep-th/0412243.}

\lref\IKV{A.~Iqbal, C.~Kozcaz, C.~Vafa,
``The Refined Topological Vertex,'' hep-th/0701156.}

\lref\CGIV{S.~Gukov, A.~Iqbal, C.~Kozcaz, C.~Vafa,
``Link homologies and the refined topological vertex,'' arXiv:0705.1368 [hep-th].}

\lref\INOV{A.~Iqbal, N.~Nekrasov, A.~Okounkov, C.~Vafa,
``Quantum foam and topological strings,''
JHEP {\bf 0804} (2008) 011, hep-th/0312022.}

\lref\AKMV{M.~Aganagic, A.~Klemm, M.~Marino, C.~Vafa,
``The topological vertex,''
Commun.\ Math.\ Phys.\  {\bf 254} (2005) 425–478, hep-th/0305132.}

\lref\GopakumarV{R.~Gopakumar, C.~Vafa,
``On the gauge theory/geometry correspondence,''
Adv.\ Theor.\ Math.\ Phys.\  {\bf 3} (1999) 1415, hep-th/9811131.}

\lref\GViii{R.~Gopakumar, C.~Vafa,
``M-theory and topological strings. I,II,'' hep-th/9809187; hep-th/9812127.}

\lref\OV{H.~Ooguri, C.~Vafa,
``Knot Invariants and Topological Strings,'' Nucl.\ Phys. {\bf B577} (2000) 419.}

\lref\HarveyM{J.~Harvey, G.~Moore, ``On the algebras of BPS states,''
Commun.\ Math.\ Phys.\  {\bf 197} (1998) 489, hep-th/9609017.}

\lref\DGR{N.~Dunfield, S.~Gukov, J.~Rasmussen,
``The Superpolynomial for Knot Homologies,'' math.GT/0505662.}

\lref\KontsevichS{M.~Kontsevich, Y.~Soibelman,
``Stability structures, motivic Donaldson-Thomas invariants and cluster transformations,''
arXiv:0811.2435.}

\lref\DenefM{F.~Denef, G.~Moore,
``Split states, entropy enigmas, holes and halos,'' hep-th/0702146.}

\lref\GMNeitzke{D.~Gaiotto, G.~Moore, A.~Neitzke,
``Four-dimensional wall-crossing via three-dimensional field theory,''
arXiv:0807.4723 [hep-th].}

\lref\JafferisM{D.~Jafferis, G.~Moore,
``Wall crossing in local Calabi Yau manifolds,'' arXiv:0810.4909 [hep-th].}

\lref\DiaconescuM{E.~Diaconescu and G.~W.~Moore,
``Crossing the Wall: Branes vs. Bundles,'' arXiv:0706.3193 [hep-th].}

\lref\JafferisC{W.~Chuang, D.~Jafferis,
``Wall Crossing of BPS States on the Conifold from Seiberg Duality and Pyramid Partitions,''
arXiv:0810.5072 [hep-th].}

\lref\Nagao{K.~Nagao,
``Derived categories of small toric Calabi-Yau 3-folds and counting invariants,''
arXiv:0809.2994.}

\lref\NagaoN{K.~Nagao, H.~Nakajima,
``Counting invariant of perverse coherent sheaves and its wall-crossing,''
arXiv:0809.2992.}

\lref\OSV{H.~Ooguri, A.~Strominger, C.~Vafa,
``Black hole attractors and the topological string,''
Phys.\ Rev.\  D {\bf 70} (2004) 106007, hep-th/0405146.}

\lref\miami{S.~Gukov, lectures at the workshop on
{\it Homological Mirror Symmetry and Related Topics,}
January 2008, University of Miami.}

\lref\inprogress{work in progress}

\lref\Szendroi{B.~Szendr\"oi,
``Non-commutative Donaldson-Thomas invariants and the conifold,''
Geom.\ Topol.\ {\bf 12} (2008) 2, arXiv:0705.3419.}

\lref\Young{B.~Young, ``Computing a pyramid partition generating function with dimer shuffling,''
J. Combin. Theory Ser. A. {\bf 116} 2 (2009), 334, arXiv:0709.3079 [math.CO].}

\lref\EKLP{N.~Elkies, G.~Kuperberg, M.~Larsen, and J.~Propp,
``Alternating-Sign Matrices and Domino Tilings II,'' J. Alg. Combin. {\bf 1} 2 (1992) 219, math/9201305.}

\lref\ORV{A.~Okounkov, N.~Reshetikhin, C.~Vafa,
``Quantum Calabi-Yau and classical crystals,'' (2003), hep-th/0309208.}

\lref\OY{H.~Ooguri, M.~Yamazaki, ``Crystal Melting and Toric Calabi-Yau Manifolds,'' (2008), hep-th/0811.2801.}

\lref\MR{S.~Mozgovoy, M.~Reineke,
``On the Noncommutative Donaldson-Thomas Invariants Arising from Brane Tilings,'' (2008), arXiv:0809.0117 [math.AG]}

\lref\Hanany{S.~Franco et al., ``Gauge theories from toric geometry and brane tilings,''
JHEP {\bf 1} (2006) 128, hep-th/0505211.\hfill\break
S.~Franco, A.~Hanany, K.~Kennaway, D.~Vegh, B.~Wecht, ``Brane Dimers and Quiver Gauge Theories,''
JHEP {\bf 1} (2006) 96, hep-th/0504110.\hfill\break
A.~Hanany, K.~Kennaway, ``Dimer models and toric diagrams,'' (2005) hep-th/0503149.}

\lref\MNOP{D.~Maulik, N.~Nekrasov, A.~Okounkov, R.~Pandharipande,
``Gromov-Witten theory and Donaldson-Thomas theory I, II''
Compos. Math. {\bf 142} (2006) 1263, 1286, math/0312059, math/0406092.}

\lref\Toda{Y.~Toda,
``Curve counting theories via stable objects I. DT/PT correspondence,''
arXiv:0902.4371.}

\lref\Reineke{M.~Reineke,
``Poisson automorphisms and quiver moduli,'' arXiv:0804.3214.}

\lref\ReinekeZ{M.~Reineke,
``Cohomology of quiver moduli, functional equations, and integrality
of Donaldson-Thomas type invariants,'' arXiv:0903.0261.}

%%%%%
\def\boxit#1{\vbox{\hrule\hbox{\vrule\kern8pt
\vbox{\hbox{\kern8pt}\hbox{\vbox{#1}}\hbox{\kern8pt}} \kern8pt\vrule}\hrule}}
\def\mathboxit#1{\vbox{\hrule\hbox{\vrule\kern8pt\vbox{\kern8pt
\hbox{$\displaystyle #1$}\kern8pt}\kern8pt\vrule}\hrule}}

%%%%%%%%%%%%%%%%%%%%%%%%%%%  FIGURES   %%%%%%%%%%%%%%%%%%%%%%%%%%%%%%%

\let\includefigures=\iftrue
\newfam\black
\includefigures
\input epsf
\def\figin{\epsfcheck\figin}\def\figins{\epsfcheck\figins}
\def\epsfcheck{\ifx\epsfbox\UnDeFiNeD
\message{(NO epsf.tex, FIGURES WILL BE IGNORED)}
\gdef\figin##1{\vskip2in}\gdef\figins##1{\hskip.5in}% blank space instead
\else\message{(FIGURES WILL BE INCLUDED)}%
\gdef\figin##1{##1}\gdef\figins##1{##1}\fi}
\def\DefWarn#1{}
\def\figinsert{\goodbreak\midinsert}
\def\ifig#1#2#3{\DefWarn#1\xdef#1{fig.~\the\figno}
\writedef{#1\leftbracket fig.\noexpand~\the\figno}%
\figinsert\figin{\centerline{#3}}\medskip\centerline{\vbox{\baselineskip12pt \advance\hsize by
-1truein\noindent\footnotefont{\bf Fig.~\the\figno:} #2}}
\bigskip\endinsert\global\advance\figno by1}
%%%
\else
\def\ifig#1#2#3{\xdef#1{fig.~\the\figno}
\writedef{#1\leftbracket fig.\noexpand~\the\figno}%
%\figinsert\figin{\centerline{#3}}\medskip\centerline{\vbox{\baselineskip12pt
%\advance\hsize by -1truein\noindent\footnotefont{\bf Fig.~\the\figno:} #2}}
%\bigskip\endinsert
\global\advance\figno by1} \fi

%%%%%%%%%%%%%%%%%%%%%%%%%%%  YOUNG TABLEUAS  %%%%%%%%%%%%%%%%%%%%%%%%%%%%%%%
%%
%%                              TABLEAUX.TEX
%%      This  macro file is for producing a ``Young Tableau'' which is
%%      an array of little squares sometimes used in mathematical physics.
%%      For instance, the command $\tableau{6 3 2}$ will produce a tableau
%%      with 6 squares in the top row, 3 in the next, and 2 in the last.
%%                                  OOOOOO
%%      This tableau will look like OOO    but made of squares instead of O's.
%%                                  OO
%%      Any number of rows may be present, each having a nonzero number of
%%      squares.
%%
%%      A tableau is math mode material, so use $ or $$ to enclose it.
%%
%%      The size and line-thickness of the little boxes are controlled by the
%%      dimension parameters --
%%              \tableauside=1.0ex              %(size)
%%              \tableaurule=0.4pt              %(line-thickness)
%%      Change them if you want.
%%
%%                                                      -- Doug Eardley 9/19/8%%
%%
\newdimen\tableauside\tableauside=1.0ex
\newdimen\tableaurule\tableaurule=0.4pt
\newdimen\tableaustep
\def\phantomhrule#1{\hbox{\vbox to0pt{\hrule height\tableaurule width#1\vss}}}
\def\phantomvrule#1{\vbox{\hbox to0pt{\vrule width\tableaurule height#1\hss}}}
\def\sqr{\vbox{%
  \phantomhrule\tableaustep
  \hbox{\phantomvrule\tableaustep\kern\tableaustep\phantomvrule\tableaustep}%
  \hbox{\vbox{\phantomhrule\tableauside}\kern-\tableaurule}}}
\def\squares#1{\hbox{\count0=#1\noindent\loop\sqr
  \advance\count0 by-1 \ifnum\count0>0\repeat}}
\def\tableau#1{\vcenter{\offinterlineskip
  \tableaustep=\tableauside\advance\tableaustep by-\tableaurule
  \kern\normallineskip\hbox
    {\kern\normallineskip\vbox
      {\gettableau#1 0 }%
     \kern\normallineskip\kern\tableaurule}%
  \kern\normallineskip\kern\tableaurule}}
\def\gettableau#1 {\ifnum#1=0\let\next=\null\else
  \squares{#1}\let\next=\gettableau\fi\next}

\tableauside=1.0ex \tableaurule=0.4pt

%%%%%%%%%%%%%%%%%%%%%  Math-style letters   %%%%%%%%%%%%%%%%%%%%%%%%
\font\cmss=cmss10 \font\cmsss=cmss10 at 7pt

\def\IB{\relax\hbox{$\inbar\kern-.3em{\rm B}$}}
\def\IC{\relax\hbox{$\inbar\kern-.3em{\rm C}$}}
\def\IQ{\relax\hbox{$\inbar\kern-.3em{\rm Q}$}}
\def\ID{\relax\hbox{$\inbar\kern-.3em{\rm D}$}}
\def\IE{\relax\hbox{$\inbar\kern-.3em{\rm E}$}}
\def\IF{\relax\hbox{$\inbar\kern-.3em{\rm F}$}}
\def\IG{\relax\hbox{$\inbar\kern-.3em{\rm G}$}}
\def\IGa{\relax\hbox{${\rm I}\kern-.18em\Gamma$}}
\def\IH{\relax{\rm I\kern-.18em H}}
\def\IK{\relax{\rm I\kern-.18em K}}
\def\IL{\relax{\rm I\kern-.18em L}}
\def\IP{\relax{\rm I\kern-.18em P}}
\def\IR{\relax{\rm I\kern-.18em R}}
\def\Z{\relax\ifmmode\mathchoice
{\hbox{\cmss Z\kern-.4em Z}}{\hbox{\cmss Z\kern-.4em Z}} {\lower.9pt\hbox{\cmsss Z\kern-.4em Z}}
{\lower1.2pt\hbox{\cmsss Z\kern-.4em Z}}\else{\cmss Z\kern-.4em Z}\fi}

\def\II{\relax{\rm I\kern-.18em I}}

\def\R{{\bf R}}
\def\C{{\bf C}}

%%%%%%%%%%%%%%%%%%%%% Calligraphic letters  %%%%%%%%%%%%%%%%%%%%%

\def\CE {{\cal E}}
\def\CF {{\cal F}}

\def\CH {{\cal H}}

\def\CM {{\cal M}}
\def\CN {{\cal N}}
\def\CO {{\cal O}}

\def\CQ {{\cal Q}}

\def\CW {{\cal W}}

\def\CZ {{\cal Z}}

%%%%%%%%%%%%%%%%%%%%%%%%%% Derivatives  %%%%%%%%%%%%%%%%%%%%%%%%

\def\p{\partial}

%%%%%%%%%%%%%%%%%%%% letters with bar %%%%%%%%%%%%%%%%%%%%%%%%%%
\def\tilde{\widetilde}
\def\hat{\widehat}
\def\bar{\overline}

%%%%%%%%%%%%%%%%%%%%%%%%%%% Math symbols %%%%%%%%%%%%%%%%%%%%%%%

\def\Tr{{\rm Tr}}

\def\p{\partial}

\def\Lie{{\rm Lie}}

\def\inbar{\,\vrule height1.5ex width.4pt depth0pt}

\def\i{{\rm Im}}

%%%%%%%%%%%%%%%%%%%   Greek letters %%%%%%%%%%%%%%%%%%%

\def\la{\lambda}

\def\bar{\overline}

\def\Tr{{\rm Tr}}

\def\IH{{\bf H}}

\def\Fl{{\CF {\kern -1.2pt \ell} }}

\def\example#1{\bgroup\narrower\footnotefont\baselineskip\footskip\bigbreak
\hrule\medskip\nobreak\noindent {\bf Example}. {\it #1\/}\par\nobreak}
\def\endexample{\medskip\nobreak\hrule\bigbreak\egroup}

\def\btimes{~{{{\lower1pt\hbox{$\square$}} \kern-7.6pt \times}}~}

\def\C{{\Bbb{C}}}

\def\rf{{\rm ref}}

%%%%%%%%%%%%%%%%%%%%%%%%%%%%%%%%%%%%%%%%%%%%%%%%%%%%%%%%%%%%%%%%%%%%%%%
%%%%%%%%%%%%%%%%%%% TITLE PAGE  %%%%%%%%%%%%%%%%%%%%%%%%%%%%%%%%

\Title{\vbox{\baselineskip11pt
\hbox{CALT-68-2725}
}}
{\vbox{
\centerline{Refined, Motivic, and Quantum}
}}
\centerline{Tudor Dimofte$^a$ and Sergei Gukov$^{a,b}$}
\medskip
\medskip
\medskip
\vskip 8pt
\centerline{$^a$ \it California Institute of Technology 452-48, Pasadena, CA 91125, USA}
\medskip
\centerline{$^b$ \it Department of Physics, University of California, Santa Barbara, CA 93106, USA}
\medskip
\medskip
\medskip
\noindent

\vskip 20pt {\bf \centerline{Abstract}} \noindent

It is well known that in string compactifications on toric Calabi-Yau
manifolds one can introduce refined BPS invariants that carry information
not only about the charge of the BPS state but also about the spin content.
In this paper we study how these invariants behave under wall crossing.
In particular, by applying a refined wall crossing formula,
we obtain the refined BPS degeneracies for the conifold in different chambers.
The result can be interpreted in terms of a new statistical model
that counts `refined' pyramid partitions;
the model provides a combinatorial realization of wall crossing and
clarifies the relation between refined pyramid partitions and the refined topological vertex.
We also compare the wall crossing behavior of the refined BPS invariants
with that of the motivic Donaldson-Thomas invariants introduced by Kontsevich-Soibelman.
In particular, we argue that, in the context of BPS state counting,
the three adjectives in the title of this paper are essentially synonymous.

\smallskip

\medskip
\Date{April 2009}

%%%%%%%%%%%%%%%%%%%%%%%%%%%%%%%%%%%%%%%%%%%%%%%%%%%%%%%%%%%%%%%%%%%%%%%

\newsec{Introduction}

This paper is devoted to the study of the space of BPS states, $\CH_{BPS}$,
in type II string compactifications on Calabi-Yau 3-folds.
In general, such compactifications lead to effective $\CN=2$ theories
in four dimensions, and, by definition, $\CH_{BPS}$ is the subspace
of the Hilbert space of an effective four-dimensional theory
that consists of one-particle states transforming
in small representations of the $d=4$, $\CN=2$ supersymmetry algebra.
The space $\CH_{BPS}$ encodes much interesting information
about the Calabi-Yau space $X$ as well as about the physics
of the four-dimensional $\CN=2$ theory, in particular providing
connections to black hole physics and topological strings \OSV.

Let us summarize some of the basic properties of the space $\CH_{BPS}$.
First, it is graded by charge sectors,
\eqn\hbps{
\CH_{BPS} = \bigoplus_{\gamma \in \Gamma} \CH_{BPS} (\gamma)\,, }
where $\Gamma$ denotes the charge lattice.
For example, in type IIA string theory\foot{As is well known,
type IIA string theory on a Calabi-Yau space $X$ is dual to
type IIB string theory on a mirror Calabi-Yau space $\tilde X$.
In what follows, we pick a duality frame corresponding to type IIA theory.}
on a Calabi-Yau 3-fold $X$
the BPS states in question are bound states of D$p$-branes with even values of $p$,
so that $\Gamma = H^{{\rm even}} (X;\Z)$ and
\eqn\mukaicharge{\matrix{
\gamma = {\rm ch} (\CE) \sqrt{\hat A (X)} & = & p^0 & + & P & + & Q & + & q_0 \cr
& \in & H^0 & \oplus & H^2 & \oplus & H^4 & \oplus & H^6 \cr
& & D6 & & D4 & & D2 & & D0 }}
is the charge vector of a D6/D4/D2/D0 bound state
(equivalently, the Mukai vector of the corresponding coherent sheaf $\CE$).
Roughly speaking, in this case $\CH_{BPS} (\gamma) \cong H^* (\CM (\gamma))$,
where $\CM (\gamma)$ denotes the moduli space of branes of charge $\gamma$.
In addition, $\CH_{BPS}$ is a representation of the rotation group $Spin(3)$
in four space-time dimensions.
This gives $\CH_{BPS}$ an extra grading, which eventually leads
to the refinement of BPS invariants considered below.
Thus, altogether, $\CH_{BPS}$ comes equipped with a $\Gamma \oplus \Z$-grading.

The space of BPS states, $\CH_{BPS}$,
depends moreover on the asymptotic boundary conditions in four space-time dimensions.
Much of its interest actually comes from the dependence on this extra data,
which includes the moduli of the Calabi-Yau 3-fold $X$ \DenefM.
Instead of working directly with $\CH_{BPS}$, it is often convenient
to consider a simpler object, the index of BPS states%
\foot{More precisely, this index is the second helicity supertrace. Following \DenefM,
in our definition of $\CH_{BPS}$ we tacitly factored out the contribution
of a universal half-hypermultiplet associated with the position in $\R^3$,
allowing us to write the index of BPS states in the simple form given here.
In $\CH_{BPS}$, a hypermultiplet counts as a state of spin zero (hence $\Omega=1$)
and a vector multiplet as spin $1/2$ (hence $\Omega=-2$).}
\eqn\oindex{
\Omega (\gamma;u) := \Tr_{\CH (\gamma;u)} (-1)^F }
that `counts' BPS states of given charge $\gamma$
and is invariant under complex structure deformations of $X$.
However, as the notation indicates, the index $\Omega (\gamma;u)$
still depends on the asymptotic value of the complexified Kahler moduli, $u = B + iJ$.
It is a piecewise constant function of $u$ that can jump
across walls of marginal stability,
where the phases of the central charges of the constituents of a bound state align.

Our main focus in this paper will be the refined BPS index,
defined as\foot{Again we factor out a contribution $y^{-1}(1-y)^2$ from a universal half-hypermultiplet.}
\eqn\orefined{
\Omega^{\rf} (\gamma; u; y) =
\sum_n (-y)^n \Omega_n^{\rf} (\gamma; u) := \Tr_{\CH (\gamma;u)} (-y)^{2J_3}\,,
}
where $J_3$ is a generator of the rotation group $Spin(3)$.
In a simplified situation where $\CH_{BPS} (\gamma; u)$ admits a description
as the cohomology of the brane moduli space $\CM (\gamma;u)$,
the BPS index $\Omega (\gamma;u)$ and its refinement $\Omega^{\rf} (\gamma;u;y)$
correspond, respectively,
to the Euler characteristic and to the Poincar\'e polynomial of $\CM (\gamma;u)$.
In particular, from the definition \orefined\ it is clear that
at $y=1$ we have $\Omega^{\rf} (\gamma; u; 1) = \Omega (\gamma;u)$.

While the refinement $\Omega^{\rf} (\gamma;u;y)$ captures useful information
about the spin content of BPS states,
it is generically not invariant under complex structure deformations of $X$.
However, if $X$ has no complex structure deformations,
then the refined BPS invariants $\Omega^{\rf} (\gamma;u;y)$ as well as
the space $\CH_{BPS} (\gamma; u)$ itself are expected to be
interesting invariants of $X$.
Thus, in certain examples the refined BPS invariants can be related
to equivariant instanton counting~\refs{\Nekrasov,\IqbalV}
or to categorification of quantum group invariants~\GSV.
In fact, in the early days of the refined BPS invariants,
the only practical way to compute them was by using one of these relations.
The situation improved significantly with the advent
of the {\it refined topological vertex}~\IKV,
which reduced the computation of refined BPS invariants
for an arbitrary non-compact toric Calabi-Yau 3-fold $X$
to a systematic combinatorial algorithm based on the counting of 3D partitions.
Note that all toric Calabi-Yau manifolds are automatically rigid
and provide an excellent laboratory for studying refined BPS invariants.
They will also be our main examples in the present paper.

In the context of local toric Calabi-Yau manifolds,
a natural object to consider is a generating function
\eqn\zddd{
Z (q,Q;u) := \sum_{{\beta \in H_2 (X;\Z) \atop n \in \Z}}
(-q)^n Q^{\beta} ~\Omega (\gamma_{\beta,n};u) }
that `counts' BPS states of D0 and D2 branes bound
to a single D6 brane.
Here, $\gamma_{\beta,n}$ is a shorthand notation
for the charge $\gamma = (1,0,-\beta,n)$ of a D6/D2/D0 system
with $n$ units of the D0-brane charge
and the D2-brane charge corresponding to a curve
in homology class $\beta \in H_2 (X;\Z)$.
In one of the chambers, the D6/D2/D0 partition function \zddd\
is the usual generating function of Donaldson-Thomas/Gopakumar-Vafa
invariants \refs{\DenefM,\JafferisM}.
Similarly, the refinement of \zddd,
\eqn\zdddref{
Z^{\rf} (q,Q,y;u) := \sum_{{\beta \in H_2 (X;\Z) \atop n \in \Z}}
(-q)^n Q^{\beta} ~\Omega^{\rf} (\gamma_{\beta,n};u;y)\,, }
carries information not only about the charges of
the D6/D2/D0 bound states but also about the spin content.
For $u$ in the DT region of the K\"ahler moduli space,
it reduces to the generating function of the usual refined BPS invariants
computed by the refined topological vertex.
One of our goals will be to study the refined partition
function \zdddref\ in other chambers and, more generally,
to understand how the refined BPS invariants $\Omega^{\rf} (\gamma;u;y)$
change across walls of marginal stability.
In particular, in the Szendr\"oi region we obtain a refinement
of the non-commutative Donaldson-Thomas partition function \Szendroi.
For example, for the resolved conifold
$X = \CO_{\Bbb{P}^1} (-1) \oplus \CO_{\Bbb{P}^1} (-1)$,
it looks like
\eqn\conifoldrfncdt{Z^{\rf}_{{\rm NCDT}}(q_1,q_2,Q) = M(q_1,q_2)^2
\prod_{i,j=1}^{\infty} \left( 1 - Q q_1^{i-{1 \over 2}} q_2^{j-{1 \over 2}} \right)
\left( 1 - Q^{-1} q_1^{i-{1 \over 2}} q_2^{j-{1 \over 2}} \right)\,, }
where instead of $q$ and $y$ we use the variables $q_1$ and $q_2$,
standard in the literature on the refined BPS invariants,
{\it cf.} \refs{\Nekrasov,\IqbalV,\IKV,\CGIV}:
\eqn\qqyy{ q_1 = qy\,,\qquad q_2 = {q \over y}\,. }

For toric Calabi-Yau 3-folds in the Szendr\"oi region of moduli space,
the generating functions of BPS states can be computed via statistical
crystal melting models \refs{\Szendroi, \MR, \OY}
that are seemingly distinct from the topological vertex formalism \refs{\AKMV,\ORV}.
Such models can be derived from quivers and brane tilings \Hanany.
In the case of the conifold, the statistical models take the form
of ``pyramid partitions,'' and it was shown \JafferisC\ that they
can be generalized to describe many chambers in the resolved conifold
moduli space outside the Donaldson-Thomas region.
In analogy with the refinement of the topological vertex, there exists a refinement of pyramid partitions
that computes (for example) the generating function \conifoldrfncdt.
A combinatorial shuffling operation on pyramid partitions corresponds
to refined wall crossing between chambers of moduli space, and,
as one approaches the Donaldson-Thomas region, the refined pyramid
partitions actually resolve into refined topological vertices.
We expect this behavior to be quite general for toric Calabi-Yau manifolds.

We observe that the wall crossing behavior of the refined BPS
invariants is very close to that of the motivic Donaldson-Thomas
invariants defined by Kontsevich and Soibelman~\KontsevichS.
In fact, the motivic Donaldson-Thomas invariants can also be
viewed as a ``refinement'' of the numerical Donaldson-Thomas invariants
that depends on the extra variable $\Bbb{L}$ (the motive of the affine line).
Equivalently, this variable can be interpreted as a ``quantum''
deformation parameter\foot{Not to be confused with the formal
variable in the generating function $Z (q,Q;u)$.}
$q$ in quantization of the complex torus ${\bf T}_{\Gamma} = \Gamma^{\vee} \otimes \C^*$,
a fact that was extensively used in \KontsevichS.
We claim that this is not an accident and the refined BPS invariants
are actually the same as the motivic BPS invariants of Kontsevich and Soibelman,
provided we identify $\Bbb{L}$ (resp. $q$) with the extra
variable $y^2$ that appears in the definition \orefined\
of the refined BPS invariants\foot{We are intentionally a little imprecise here.
We intend to say that the deformation by $y$ on the `refined' side
corresponds to the deformation by $-q^{1/2}$ on the `motivic' side.
While in general there may be a non-trivial map between these
two deformations, as we explain below in section~3
these deformations actually agree to the leading order.}:
\eqn\rmq{\matrix{
\underline{\rm Refined} && \underline{\rm Motivic} && \underline{\rm Quantum} \cr
y & \longleftrightarrow & \Bbb{L}^{1/2}
& \longleftrightarrow & -q^{1/2} } }
Further evidence for this identification comes from the connection
with homological invariants of knots and 3-manifolds that will be discussed
elsewhere \inprogress.
Note that the semi-classical limit corresponds to $q^{1/2} \to -1$ (resp. $y \to 1$).

As is well know, in string compactification on a Calabi-Yau
3-fold, the variable $y$ can be interpreted as a graviphoton
background in four space-time dimensions.
Therefore, we propose the following

\medskip\noindent
{\bf Conjecture:} {\it In string theory on a Calabi-Yau space $X$,
the $q$-deformation of \KontsevichS\ corresponds to turning on
a graviphoton background (a.k.a. $\Omega$-background) on $\R^4$.
In particular, we have}
\eqn\motrf{\mathboxit{
\Omega^{\rf} (\gamma;u;y) = \Omega^{{\rm mot}} (\gamma;u)
}}
{\it with the appropriate identification of variables \rmq.}

Although in practice it is easier to work with numerical invariants
such as $\Omega (\gamma;u)$ or $\Omega^{\rf} (\gamma;u;y)$,
we should emphasize that the ultimate goal is to ``categorify''
the wall crossing formulae for the numerical BPS invariants
and to explore the properties of the space $\CH_{BPS}$ itself.
We believe this should lead to a rich mathematical structure.
(See \refs{\HarveyM,\DGR} for earlier work where
the homological algebra of BPS states played an important role.)
In fact, it seems that the homological algebra of $\CH_{BPS}$
is unavoidable if one tries to study the refined BPS invariants.
For example, there can be walls --- we call them ``invisible walls'' ---
where the ordinary index $\Omega (\gamma;u)$ doesn't change,
but $\Omega^{\rf} (\gamma;u;y)$ jumps. The basic
mechanism for how this happens can be understood in
a simple situation where $\CH_{BPS}$ admits a description
as cohomology of the brane moduli space $\CM$.
Then, as $u$ crosses an invisible wall, $\CM$ can develop
a singularity and undergo a topology-changing transition,
so that the Poincar\'e polynomial of $\CM$ changes
while the Euler characteristic does not.

These invisible walls have been observed in \DGR\ in certain
instances of BPS state counting and will be discussed further in \inprogress.
One of their most interesting properties is that the best way
to describe the change in the spectrum of refined BPS invariants across them
is by observing that in K\"ahler moduli space these walls are located
where two BPS states in short multiplets can combine into a long multiplet.
Therefore, the states that disappear from the spectrum
are trivial in cohomology of the BRST differential
\eqn\qdifferential{
{\CQ} : \CH^n_{BPS} \to \CH^{n+1}_{BPS}\,, }
where $n=2J_3$ denotes the $\Z$-grading of $\CH_{BPS}$ by spin.
(Note that $\CQ$ changes spin by~$\half$.)
At the level of the generating function \zdddref, it means that
across each invisible wall the change in $Z^{\rf} (q,Q,y;u)$
includes an elementary factor $(1-y)$.

\bigskip {\noindent {\it {Organization of the Paper}}}

In section 2, following \refs{\DenefM,\JafferisM}
(see also \refs{\JafferisC,\DiaconescuM,\NagaoN}),
we briefly review some relevant facts about the wall crossing behavior
of the BPS index $\Omega (\gamma;u)$ and how it can be used
to compute the D6/D2/D0 partition function in concrete examples,
such as the resolved conifold.
We then generalize this discussion to refined BPS invariants
and compute $Z^{\rf} (q,Q,y;u)$ for the conifold in different chambers,
the refined non-commutative Donaldson-Thomas partition
function \conifoldrfncdt\ being a special case.
In section 3, we compare the wall crossing behavior of the refined BPS invariants
with that of the motivic Donaldson-Thomas invariants introduced by Kontsevich-Soibelman \KontsevichS.
In particular, one way to see the proposed identification \motrf\
is to deduce the primitive (or semi-primitive) wall crossing formula
for the refined BPS invariants from the motivic wall crossing formula
of Kontsevich and Soibelman. As we shall see in section 3,
in this derivation, one is naturally led to the identification
of the ``deformation'' parameters in eq. \rmq.
Finally, in section 4, we interpret the results of section 2
in terms of the new statistical model that counts refined pyramid partitions.
In both refined and unrefined cases, we explain how a shuffling operation
on pyramid partitions provides a combinatorial realization of wall crossing,
and we clarify the relation between pyramid partitions
and the refined and unrefined topological vertices.

%%%%%%%%%%%%%%%%%%%%%%%%%%%%%%%%%%%%%%%%%%%%%%%%%%%%%%%%%%%%%%%%%%%%%%%

\newsec{Refined Wall Crossing}

The main goal of this section is to study the wall crossing behavior
of the refined BPS invariants $\Omega^{\rf} (\gamma;u;y)$ and,
in particular, to compute the refined D6/D2/D0 partition
functions $Z^{\rf} (q,Q,y;u)$ in different chambers.

\bigskip\noindent {\it {Walls and Chambers}}

We are interested in walls of marginal stability for decays
$$
\gamma \to \gamma_1 + \gamma_2\,.
$$
In the K\"ahler moduli space, such decays take place at the points $u$
where the central charges $\CZ (\gamma_1;u)$ and $\CZ (\gamma_2;u)$
of the constituents align;
the corresponding walls will be denoted as $\CW (\gamma_1,\gamma_2)$:
\eqn\wallpos{
\CW (\gamma_1,\gamma_2) = \{ ~u~\vert~ \CZ (\gamma_1;u) = \lambda \CZ (\gamma_2;u)
~~~{\rm for~some~} \la \in \IR_+ \}\,. }
Note, in particular, that all walls $\CW (N_1 \gamma_1, N_2 \gamma_2)$ coincide,
$$
\CW (\gamma_1, \gamma_2) = \CW (N_1 \gamma_1, N_2 \gamma_2)\,,
\quad\quad\quad
N_1, N_2 \in \Z_+\,.
$$
The bound state with charge $\gamma = \gamma_1 + \gamma_2$
is stable on the side of the wall $\CW (\gamma_1,\gamma_2)$ where
\eqn\stableineq{
\langle \gamma_1 , \gamma_2 \rangle \,\i \CZ (\gamma_1;u) \bar{\CZ (\gamma_2;u)} > 0 \,.}
This condition, as well as the position of the wall \wallpos,
has an elegant interpretation in supergravity,
if all the charges $\gamma_i$ are large \DenefM.
For example, in the case of a 2-center bound state,
the separation of the two constituents in $\R^3$
is determined by their charges,
$$
R_{12} = {1 \over 2} \langle \gamma_1 , \gamma_2 \rangle
{|\CZ_1 + \CZ_2|_{\infty} \over \i (\CZ_1 \bar \CZ_2)_{\infty}\,}\,,
$$
where $\CZ (\gamma)$ is the central charge function.
In particular, we see that the condition \stableineq\ is necessary
for the distance $R_{12}$ to be positive, and $R_{12}$ diverges when $u$
approaches a wall of marginal stability $\CW (\gamma_1,\gamma_2)$.
For future reference, we also note that the 2-center bound state
carries angular momentum \DenefM:
\eqn\spinonetwo{
J_{12} = {1 \over 2} \left( \vert \langle \gamma_1 , \gamma_2 \rangle \vert - 1 \right)\,. }

In the context of counting BPS states of D0 and D2 branes bound to
a single D6 brane on a toric Calabi-Yau manifold $X$,
one encounters walls of marginal stability with $\gamma_1 = (1,0,- m',n')$
and a ``halo'' of particles or ``fragments'' of charge $\gamma_2 = (0,0,- m_h, n_h)$.
Following \refs{\JafferisM,\JafferisC}, we denote such walls as $\CW_{n_h}^{m_h}$.
For example, for the resolved conifold
$X = \CO_{\Bbb{P}^1} (-1) \oplus \CO_{\Bbb{P}^1} (-1)$,
the non-vanishing Gopakumar-Vafa invariants \GViii,
\eqn\gvconifold{\eqalign{
& \Omega (\gamma = (0,0,\pm 1,n)) = 1
\quad\quad\quad n \in \Z \cr
& \Omega (\gamma = (0,0,0,n)) = -2
\quad\quad\quad n \ne 0\,,
}}
imply that the only walls are
\eqn\conifoldwalls{\eqalign{
\CW_n^{1} &: \quad\quad D2/D0 {\rm ~fragments} \cr
\CW_n^{-1} &: \quad\quad \bar{D2}/D0 {\rm ~fragments} \cr
\CW_n^{0} &: \quad\quad D0 {\rm ~fragments}\,. }}
This leads to the following picture of walls and chambers
(in the one-dimensional space parameterized by $\varphi$)~\JafferisM:

\ifig\wallsfig{The picture of walls and chambers for the resolved conifold proposed in~\JafferisM.}
{\epsfxsize3.0in\epsfbox{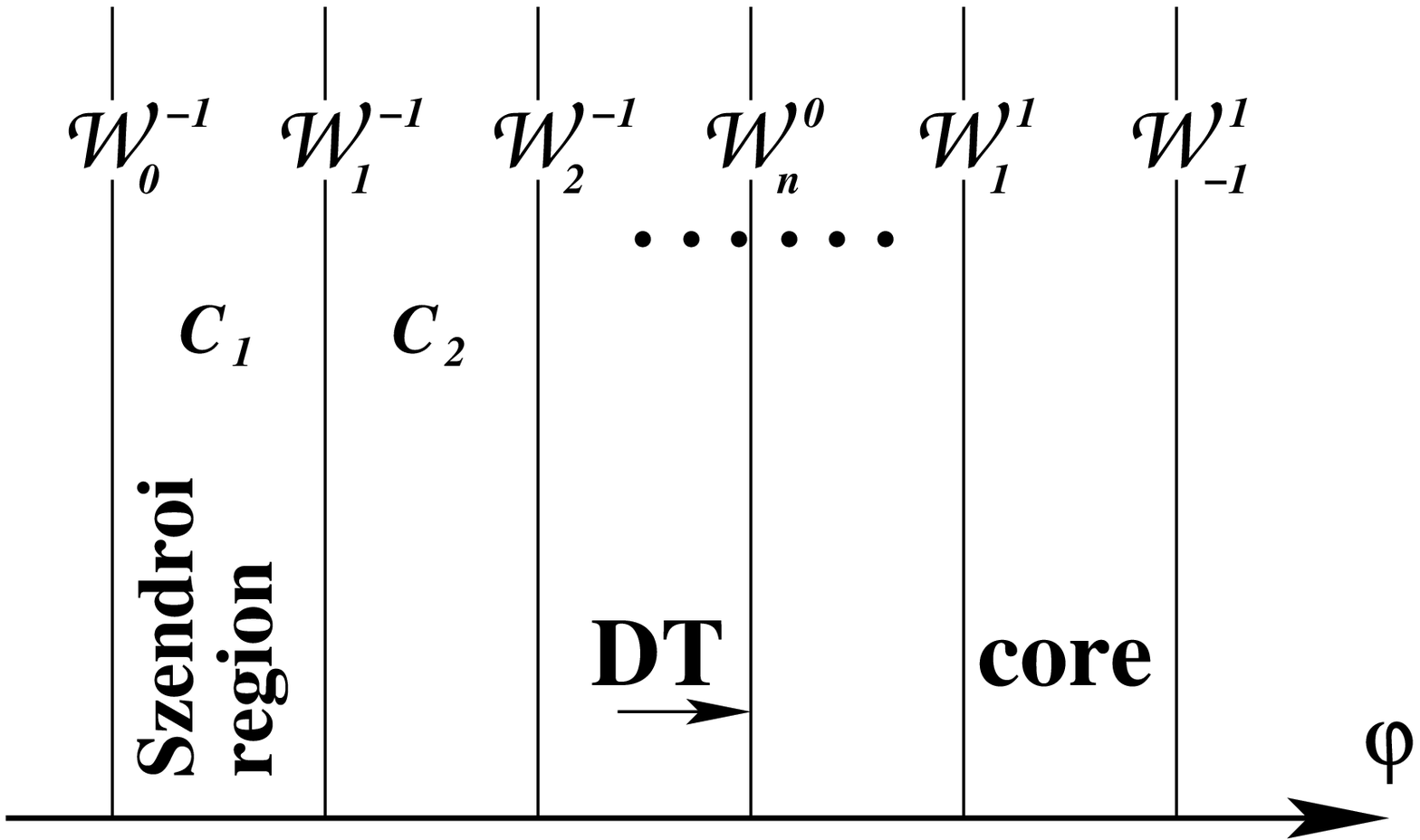}}

\noindent
We follow the conventions of \refs{\JafferisM,\JafferisC},
and denote the resulting chambers for the resolved conifold as
\eqn\conifoldchambers{\eqalign{
C_n & = [\CW_n^{-1} \CW_{n-1}^{-1}] \cr
 \tilde{C}_n & = [\CW_{n-1}^{1} \CW_{n}^{1}]\,. }}
%

%%%%%%%%%%%%%%%%%%%%%%%%%%%%%%%%%%%%%%%%%%%%%%%%%%%%%%%%%%%%%%%%%%%%%%%

%\subsec{Wall Crossing Formulae and Refinements}
\bigskip\noindent {\it {Wall Crossing Formulae and Refinements}}

If both $\gamma_1$ and $\gamma_2$ are primitive,
the states lost from $\CH_{BPS} (\gamma; u)$ are \DenefM
\eqn\hbpsprimitive{
\Delta \CH_{BPS} = (J_{12}) \otimes \CH (\gamma_1; u_{ms}) \otimes \CH (\gamma_2; u_{ms})\,, }
where $J_{12}$ is the spin of the bound state \spinonetwo.
The corresponding change of the BPS index \oindex\ is given by
\eqn\primitivewcf{
\Delta \Omega (\gamma;u) = (-1)^{\langle \gamma_1, \gamma_2 \rangle - 1}
|\langle \gamma_1, \gamma_2 \rangle| \Omega (\gamma_1; u_{ms}) \Omega (\gamma_2; u_{ms})\,. }
This is the so-called primitive wall crossing formula.
It has a very explicit, but lengthier, generalization
in the case when one of the charges is not primitive.
It is then convenient to arrange the answer
in a generating function that allows one to describe the change
of the spectrum of BPS states as a sum over Fock spaces \DenefM:
\eqn\hbpssemiprimitive{
\bigoplus_{N_2} x^{N_2} \Delta \CH \vert_{\gamma \to \gamma_1 + N_2 \gamma_2} =
\CH (\gamma_1) \bigotimes_k \CF \left( x^k (J_{\gamma_1, k \gamma_2}) \otimes \CH (k \gamma_2) \right)\,, }
so that
\eqn\semiprimitivewcf{
\Omega (\gamma_1) + \sum_{N \ge 1} \Delta \Omega (\gamma_1 + N \gamma_2) x^N
= \Omega (\gamma_1) \prod_{k \ge 1}
(1 - (-1)^{k \langle \gamma_1, \gamma_2 \rangle} x^k)^{k |\langle \gamma_1, \gamma_2 \rangle| \Omega (k \gamma_2)}\,. }
This is the semi-primitive wall crossing formula.
Note that ${\p \over \p x} (\ldots) \vert_{x=0}$
gives the primitive wall crossing formula \primitivewcf.

In general, if neither $\gamma_1$ nor $\gamma_2$ is primitive,
the change of the BPS index across a wall of marginal stability
is given by the Kontsevich-Soibelman
wall crossing formula~\refs{\KontsevichS,\GMNeitzke}
that will be discussed in section 3 below.
We note in passing that if one is interested only in counting
D6/D2/D0 bound states on toric Calabi-Yau manifolds,
then the primitive \primitivewcf\
and semi-primitive \semiprimitivewcf\ wall crossing formulae
and their refinements (which we discuss momentarily)
are sufficient to determine the D6/D2/D0 partition functions.

Now, let us generalize this discussion to the refined BPS
invariants that, along with the charges of BPS states, also
carry information about the spin content.
Again, starting with the simplest case when both charges
$\gamma_1$ and $\gamma_2$ are primitive,
the change of $\Omega^{\rf} (\gamma;u;y)$ defined in \orefined\
is given by
\eqn\rfprim{\eqalign{
& \Delta \Omega^{\rf} (\gamma;u;y) = 
{(-y)^{\langle\gamma_1,\gamma_2\rangle}
   -(-y)^{-\langle\gamma_1,\gamma_2\rangle} \over
(-y)-(-y)^{-1}}
\Omega^{\rf} (\gamma_1;u;y) \Omega^{\rf} (\gamma_2;u;y) \cr
& ~~= \big( (-y)^{- \langle \gamma_1 , \gamma_2 \rangle + 1} + (-y)^{- \langle \gamma_1 , \gamma_2 \rangle + 3}
+ \ldots + (-y)^{\langle \gamma_1 , \gamma_2 \rangle - 1} \big)
\Omega^{\rf} (\gamma_1;u;y) \Omega^{\rf} (\gamma_2;u;y)\,.  }}
This is the refined version of the primitive wall crossing
formula~\primitivewcf, whose explicit form appeared in~\DiaconescuM.
Indeed, note that setting $y=1$ gives the primitive wall crossing formula~\primitivewcf.

For our purposes, we need a refinement of the semi-primitive
wall crossing formula~\semiprimitivewcf. It takes the form
\eqn\rfsprim{\eqalign{
\Omega^{\rf} (\gamma_1) &+ \sum_{N \ge 1} \Delta \Omega^{\rf} (\gamma_1 + N \gamma_2) x^N \cr
&=\; \Omega^{\rf} (\gamma_1) \prod_{k \ge 1} \prod_{j = 1}^{k |\langle \gamma_1, \gamma_2 \rangle|} \prod_n
(1 + x^k y^n (-y)^{2j - k |\langle \gamma_1, \gamma_2 \rangle|-1})^{(-1)^n \Omega_n^{\rf} (k \gamma_2)}\,. }}
where $\Omega_n^{\rf} (\gamma)$ are the coefficients of the refined BPS index \orefined.
This refinement of \semiprimitivewcf\ describes the change of the spectrum
\hbpssemiprimitive\ while keeping track of the spin of BPS states,
and satisfies two obvious requirements:

\item{$i)$} at $y=1$ it reduces to the ordinary semi-primitive wall crossing formula \semiprimitivewcf;

\item{$ii)$} ${\p \over \p x} (\ldots) \vert_{x=0}$ gives the refined primitive wall crossing formula \rfprim.

\noindent
In addition, the refinement \rfsprim\ leads to the expected results in simple special examples.

%%%%%%%%%%%%%%%%%%%%%%%%%%%%%%%%%%%%%%%%%%%%%%%%%%%%%%%%%%%%%%%%%%%%%%%
\bigskip\bigskip % these two are introdued to push the title below to the next page

\bigskip\noindent {\it {Example: Resolved Conifold}}

Now, with the simple example of the resolved conifold,
$X = \CO_{\Bbb{P}^1} (-1) \oplus \CO_{\Bbb{P}^1} (-1)$, let us illustrate how one can use these wall crossing formulae to obtain
the (refined) D6/D2/D0 partition functions in different chambers \conifoldchambers.
This discussion will be mirrored by the pyramid partition approach of Section 4.
Before we present refined BPS invariants, let us briefly review
the results of \refs{\JafferisM,\Szendroi} for the ordinary
partition functions $Z (q,Q;u)$.
Starting in the core region (see \wallsfig) where $Z(q,Q;C_{{\rm core}})=1$
and applying the wall crossing formulae, we obtain \JafferisM
\eqn\zcnt{ Z (q,Q; \tilde C_{n+1}) = \prod_{j=1}^n (1 - q^j Q)^j\,, }
so that in the limit $n\to\infty$ we recover the {\it reduced} Donaldson-Thomas
partition function
\eqn\zdtredconifold{
\lim_{n \to \infty} Z (q,Q; \tilde C_{n})
= \prod_{j>0} (1 - q^j Q)^j = Z_{DT}' (q,Q)\,. }
Similarly, in the chamber $C_{n+1} = [\CW_{n+1}^{-1} \CW_{n}^{-1}]$,
\eqn\zcn{Z (q,Q; C_{n+1}) = M(q)^2 \prod_{j > 0} (1 - q^j Q)^{j}
\prod_{k > n} (1 - q^k Q^{-1})^k\,. }
The factor of $M(q)^2 = \prod_{j=1}^\infty (1-q^j)^{-2j}$ comes from crossing the D6/D0 wall $\CW_{n}^{0}$.
In the limit $n\rightarrow\infty$, we now recover the full Donaldson-Thomas partition function,
\eqn\zdtconifold{ \lim_{n \to \infty} Z (q,Q; C_{n})
= M(q)^2 \prod_{j>0} (1 - q^j Q)^j = Z_{DT} (q,Q)\,. }

Now, starting with $Z^{\rf} (q_1,q_2,Q; \tilde C_{{\rm core}})=1$
and applying the refined semi-primitive wall crossing formula \rfsprim\
we obtain the refined D6/D2/D0 partition functions in all chambers $\tilde C_n$,
\eqn\zcnref{
Z^{\rf} (q_1,q_2,Q; \tilde C_{n+1}) = \prod_{j=1}^n \prod_{i=1}^{n-j+1}
\left( 1 - Q q_1^{i-{1 \over 2}} q_2^{j-{1 \over 2}} \right)\,, }
so that in the limit we recover a refinement of the reduced Donaldson-Thomas
partition function:
\eqn\zdtredconifoldref{
\lim_{n \to \infty} Z^{\rf} (q_1,q_2,Q; \tilde C_{n})
= \prod_{i,j=1}^{\infty} \left( 1 - Q q_1^{i-{1 \over 2}} q_2^{j-{1 \over 2}} \right)\,. }
In the chamber $C_{n+1} = [\CW_{n+1}^{-1} \CW_{n}^{-1}]$, we find
\eqn\zcntref{
Z^{\rf} (q_1,q_2,Q; C_{n+1}) = M(q_1,q_2)^2
\prod_{i,j=1}^{\infty} \left( 1 - Q q_1^{i-{1 \over 2}} q_2^{j-{1 \over 2}} \right)
\prod_{k+l > n}^{\infty} \left( 1 - Q^{-1} q_1^{k-{1 \over 2}} q_2^{l-{1 \over 2}} \right)\,. }
As in the unrefined case,
the factor of $M(q_1,q_2)^2$ comes from crossing the D6/D0 wall.
In general, when studying refined BPS invariants one encounters a family of refinements 
$M_{\delta} (q_1,q_2) = \prod_{i,j=1}^\infty(1-q_1^{i-{1\over 2}+{\delta \over 2}}q_2^{j-{1\over 2}-{\delta \over 2}})^{-1}$
of the MacMahon function $M(q)$, with different values of $\delta$.
For example, the refinement used in \IKV\ corresponds to $\delta = -1$.
In the classical limit $y \to 1$ (corresponding to $q_1 = q_2 = q$)
all of these refinements reduce to the ordinary MacMahon function $M(q)$,
while in the ``opposite'' limit $y \to -1$ they specialize
to $M(-q)$, which describes the contribution of the 0-dimensional
subschemes to the $\hat{{\rm DT}}$-invariants of~\Toda.
For our purposes in the present paper, it is convenient to choose a symmetric
normalization in \zcntref, so that $M(q_1,q_2) = M_{\delta} (q_1,q_2)$ with $\delta=0$.

In the limit $n \to \infty$ we obtain a refinement of
the ordinary Donaldson-Thomas partition function \zdtconifold,
\eqn\zdtconifoldref{
\lim_{n \to \infty} Z^{\rf} (q_1,q_2,Q; C_{n}) = M(q_1,q_2)^2
\prod_{i,j=1}^{\infty} \left( 1 - Q q_1^{i-{1 \over 2}} q_2^{j-{1 \over 2}} \right)\,, }
and
\eqn\zncdtrefconifold{
Z^{\rf} (q_1,q_2,Q; C_{1}) = M(q_1,q_2)^2
\prod_{i,j=1}^{\infty} \left( 1 - Q q_1^{i-{1 \over 2}} q_2^{j-{1 \over 2}} \right)
\left( 1 - Q^{-1} q_1^{i-{1 \over 2}} q_2^{j-{1 \over 2}} \right)\, }
is a refinement of the non-commutative Donaldson-Thomas partition function $Z(q,Q; C_{1})$.

%%%%%%%%%%%%%%%%%%%%%%%%%%%%%%%%%%%%%%%%%%%%%%%%%%%%%%%%%%%%%%%%%%%%%%%

\newsec{Refined = Motivic}

The unrefined wall crossing formula of Kontsevich and Soibelman \KontsevichS\
generalizes the primitive \primitivewcf\ and semiprimitive \semiprimitivewcf\ cases discussed above.
It encodes the degeneracies of BPS states in a given chamber in terms of a non-commuting
product of symplectomorphisms acting on the complexified charge lattice. Specifically, let
\eqn\classT{{\bf T}_{\Gamma} = \Gamma^{\vee} \otimes \C^*\, }
be an $r$-dimensional complex torus, where $r$ is the rank of $\Gamma$,
and define functions $X_\gamma$ corresponding to any $\gamma \in \Gamma$,
such that $X_{\gamma} X_{\gamma'} = X_{\gamma + \gamma'}$.
Given a basis $\{ \gamma_i \}$ of $\Gamma$ and corresponding coordinates $X_i$ on ${\bf T}_{\Gamma}$,
one endows ${\bf T}_{\Gamma}$ with a symplectic structure
$\omega = {1 \over 2} \langle \gamma_i, \gamma_j \rangle^{-1} d \log X_i \wedge d \log X_j$
and defines symplectomorphisms\foot{See \KontsevichS\ and also \GMNeitzke\
(which we follow here in the unrefined case)
for a more precise description of these symplectomorphisms.}
\eqn\classU{ U_\gamma \,:\, X_{\gamma'} \to X_{\gamma'}(1\pm X_\gamma)^{\langle \gamma',\gamma\rangle}\,. }
The statement of wall crossing is that the product over all
states that become aligned at a wall of marginal stability
\eqn\KSwcf{
A = \prod_\gamma^\curvearrowright A_{\gamma} (u)
:= \prod_\gamma^\curvearrowright U_\gamma^{\Omega(\gamma;u)}\,, }
taken in order of increasing phase of the central charge, $\CZ (\gamma)$,
is the same on both sides of the wall.
In other words, going from $u=u_+$ on one side of the wall to $u=u_-$ on the other,
both the BPS indices and the ordering will change but the overall product will remain the same:
\eqn\kswcfclass{
\prod_{\gamma}^{{\curvearrowright}} U_{\gamma}^{\Omega (\gamma,u_+)}
= \prod_{\gamma}^{{\curvearrowright}} U_{\gamma}^{\Omega (\gamma,u_-)} \,. }

Geometrically, this formula arises as a specialization
(namely, by taking the Euler characteristic)
of a much more general, `motivic' wall crossing formula \KontsevichS\
that is reminiscent of the wall crossing formulae
for the refined BPS invariants discussed in the previous section.
Just as refined BPS invariants depend on an extra variable $y$,
motivic BPS invariants involve a formal variable $\Bbb{L}^{1/2}$
(where $\Bbb{L}$ is the motive of the affine line).
Both refinements reduce to the ordinary BPS invariants
in the corresponding limits $y \to 1$ and $\Bbb{L}^{1/2} \to 1$.
One may therefore hope that motivic BPS invariants are precisely
the refined BPS invariants of physics.

There are several ways to approach such a conjecture.
First, one can attempt to make a direct comparison of refined
and motivic BPS invariants in concrete examples,
several of which are attainable and will be discussed in~\inprogress.
Second, the behavior of BPS invariants can be investigated indirectly
by means of wall crossing formulae. This is our goal in the present paper.
Using the Serre functor as in~\KontsevichS,
we pass to a ``quantum'' version of motivic BPS invariants and the wall crossing formula,
and show that it agrees with~\rfprim.
Finally, and more physically, one can try to derive the motivic wall crossing formula
via three-dimensional gauge in the presence of a graviphoton background,
along the lines of~\GMNeitzke.

The motivic DT invariants defined by Kontsevich and Soibelman
are elements of quantum tori over a version of the Grothendieck ring
of varieties \KontsevichS.
The quantum torus in question is simply the quantization of \classT.
It comprises an associative algebra generated by $\hat e_{\gamma}$, $\gamma \in \Gamma$,
such that\foot{In the case of the motivic quantum torus
this relation looks like $\hat e_{\gamma_1} \hat e_{\gamma_2}
= \Bbb{L}^{\half \langle \gamma_1, \gamma_2 \rangle} ~\hat e_{\gamma_1 + \gamma_2}$.}
\eqn\eeq{
\hat e_{\gamma_1} \hat e_{\gamma_2}
= q^{\half \langle \gamma_1, \gamma_2 \rangle} ~\hat e_{\gamma_1 + \gamma_2} }
and $\hat e_0 = 1$.
In particular, the generators obey the following commutation relations:
$$
[\hat e_{\gamma_1} , \hat e_{\gamma_2} ]
= \left( q^{\half \langle \gamma_1, \gamma_2 \rangle}
- q^{-\half \langle \gamma_1, \gamma_2 \rangle} \right) \hat e_{\gamma_1 + \gamma_2}\,.
$$
In the classical limit, as $q^{1/2} \to -1$, one finds
$$
\lim_{q^{1/2} \to -1} (q-1)^{-1} \left( q^{\half \langle \gamma_1, \gamma_2 \rangle}
- q^{-\half \langle \gamma_1, \gamma_2 \rangle} \right)
= (-1)^{\langle \gamma_1, \gamma_2 \rangle} \langle \gamma_1, \gamma_2 \rangle\,,
$$
so that
\eqn\Lie{[e_{\gamma_1}, e_{\gamma_2} ]
= (-1)^{\langle \gamma_1, \gamma_2 \rangle} \langle \gamma_1, \gamma_2 \rangle
e_{\gamma_1 + \gamma_2}\,,}
where
$$
e_{\gamma} := \lim_{q^{1/2} \to -1} {\hat e_{\gamma} \over q-1}\,.
$$
The Lie algebra \Lie\ acts on the classical torus \classT,
and generates symplectomorphisms via $U_\gamma = \exp\sum_{n=1}^\infty {1\over n^2}e_{n\gamma}$.

Following Kontsevich and Soibelman \KontsevichS,
let us introduce the ``quantum dilogarithm'' function
\eqn\edilog{
{\bf E} (x) := \sum_{n=0}^{\infty} {q^{n^2/2} \over (q^n - 1) \ldots (q^n - q^{n-1})} x^n\,. }
It is easy to verify that in the classical limit $q^{1/2} \to -1$
this function has the asymptotic expansion
$$
{\bf E} (x) =
\exp \left( - {1 \over 2\hbar} {\rm Li}_2 (x) + {x \hbar \over 12 (1-x)} + \ldots \right)
$$
where $q^{1/2} = - e^{\hbar}$
and ${\rm Li}_2 (x) = \sum_{n=1}^{\infty} {x^n \over n^2}$ is the Euler dilogarithm.
Moreover, the function ${\bf E} (x)$ obeys the ``pentagon'' identity
\eqn\eeeee{
{\bf E} (x_1) {\bf E} (x_2) = {\bf E} (x_2) {\bf E} (x_{12}) {\bf E} (x_1)\,, }
where $x_1 x_2 = q x_2 x_1$ and $x_{12} = q^{-1/2} x_1 x_2 = q^{1/2} x_2 x_1$.
This pentagon identity is the basic example of the motivic wall crossing formula
in a simple case of two primitive charges, $\gamma_1$ and $\gamma_2$,
which obey $\langle \gamma_1, \gamma_2 \rangle = 1$,
\eqn\aaaaa{
A_{1,0}^{{\rm mot}} \cdot A_{0,1}^{{\rm mot}}
= A_{0,1}^{{\rm mot}} \cdot A_{1,1}^{{\rm mot}} \cdot A_{1,0}^{{\rm mot}} \,. }
Here, $A_{m,n}^{{\rm mot}} : = {\bf E} (\hat e_{n \gamma_1 + m \gamma_2})$
are quantum analogs of the classical symplectomorphisms $U_{\gamma}^{\Omega (\gamma)}$.
Acting by conjugation, these operators generate ``symplectomorphisms''
of the quantum torus (resp. motivic quantum torus).

In general, the Kontsevich-Soibelman wall crossing formula for
motivic DT invariants $\Omega^{{\rm mot}} (\gamma;u)$
says that the product of the quantum symplectomorphisms
in a given sector does not change under wall crossing
(as long as no BPS rays leave the sector):
\eqn\kswcfquant{
\prod_{\gamma}^{{\curvearrowright}} A_{\gamma}^{{\rm mot}} (u_+)
= \prod_{\gamma}^{{\curvearrowright}} A_{\gamma}^{{\rm mot}} (u_-) \,. }
For simplicity, we assume here that each ray in this sector is generated
by a single BPS charge $\gamma$.
In this case, we have ({\it cf.} \KSwcf\ above and sec. 6.2 of \KontsevichS)
\eqn\aquant{
A_{\gamma}^{{\rm mot}} (u)
= 1 + {q^{1/2} \Omega^{{\rm mot}} (\gamma; u) \over q-1} \hat e_{\gamma} + \ldots\,. }
In particular, suppose $\gamma_1$ and $\gamma_2$ are primitive
and consider a narrow sector containing $\CZ (\gamma_1 + \gamma_2)$.
Then the motivic wall crossing formula \kswcfquant\ looks like
\eqn\aaabaaa{
A_{\gamma_1}^{{\rm mot}} (u_+)
A_{\gamma_1 + \gamma_2}^{{\rm mot}} (u_+)
A_{\gamma_2}^{{\rm mot}} (u_+)
=
A_{\gamma_2}^{{\rm mot}} (u_-)
A_{\gamma_1 + \gamma_2}^{{\rm mot}} (u_-)
A_{\gamma_1}^{{\rm mot}} (u_-) \,.
}
To use this formula in practice, notice that the algebra generated by
$\hat e_{n \gamma_1 + m \gamma_2}$ is filtered, {\it cf.} \GMNeitzke.
The lowest degree of filtration contains the Heisenberg subalgebra
\eqn\eecomm{
[\hat e_{\gamma_1} , \hat e_{\gamma_2} ]
= \left( q^{\half \langle \gamma_1, \gamma_2 \rangle}
- q^{-\half \langle \gamma_1, \gamma_2 \rangle} \right) \hat e_{\gamma_1 + \gamma_2}\,, }
with a central element $\hat e_{\gamma_1 + \gamma_2}$.
Then, assuming that $\Omega^{{\rm mot}} (\gamma_i,u_+) = \Omega^{{\rm mot}} (\gamma_i,u_-)$ for $i=1,2$
and keeping track of the coefficients of $\hat e_{\gamma_1 + \gamma_2}$ in \aaabaaa,
we can derive
$$
\Delta \Omega^{{\rm mot}} (\gamma_1 + \gamma_2)
= \langle \gamma_1 , \gamma_2 \rangle_q \cdot \Omega^{{\rm mot}} (\gamma_1) \Omega^{{\rm mot}} (\gamma_2)\,,
$$
where
$$
\langle \gamma_1 , \gamma_2 \rangle_q
= {q^{\half \langle \gamma_1, \gamma_2 \rangle}
- q^{-\half \langle \gamma_1, \gamma_2 \rangle} \over q^{\half} - q^{- \half}}\,.
$$
This is equivalent to the refined version \rfprim\ of the primitive wall crossing formula,
provided that we identify $y \leftrightarrow - q^{1/2}$.

%%%%%%%%%%%%%%%%%%%%%%%%%%%%%%%%%%%%%%%%%%%%%%%%%%%%%%%%%%%%%%%%%%%%%%%

\newsec{Refined Pyramid Partitions and the Topological Vertex}

In this final section, we interpret refined BPS invariants 
in terms of statistical models of melting crystals.
In the case of Donaldson-Thomas theory for non-compact toric Calabi-Yau 3-folds,
this was done in \ORV\ (unrefined) and \IKV\ (refined)
using the topological vertex formalism.
For other chambers of moduli space, however, rather different models are needed.
These models are built from Calabi-Yau quivers with superpotentials;
crystal partitions encode information about the representations of the quivers,
which in turn are related to configurations of BPS branes.
(See, for example, \refs{\Szendroi,\MR,\OY} for explicit examples
of this relation in the Szendr\"oi region of moduli space.)
Both representations of the quivers and BPS branes
depend on a stability condition (a choice of K\"ahler moduli),
which corresponds to a boundary condition for the crystal model.

In the case of the conifold, which shall be our main example,
the crystal models that arise are called ``pyramid partitions.''
They essentially describe representations of the Klebanov-Witten quiver.
By changing the boundary conditions for these crystals,
unrefined invariants in all the chambers $C_n$ and $\tilde{C}_n$
that were described in Section 2 can be obtained \JafferisC.
It turns out that by modifying the weights atoms in the crystals --- more or less
splitting weights $q_1$ and $q_2$ across a diagonal --- it is also possible
to obtain refined invariants in all chambers.
Part of our goal in this section is to describe how this is achieved.

More interestingly, we find that a combinatorial transformation between
partitions with different boundary conditions called dimer shuffling \Young\
corresponds very naturally to (refined or unrefined) wall crossing.
Moreover, in the limit $n\to\infty$ the boundary conditions for pyramid partitions
in chambers $C_n$ become such that the model resolves into a pair of
(refined or unrefined) topological vertices.
This is exactly what one should expect after crossing the infinite number
of walls to the Donaldson-Thomas chamber of moduli space for the resolved conifold.
As the topological vertex provides a universal construction for the BPS partition
function of toric Calabi-Yau 3-folds, we believe that this behavior should be quite general:
a quiver-related crystal model should always resolve
into a network of topological vertices as one approaches the DT chamber.

We first review the pyramid partition models for the unrefined BPS invariants,
and then proceed to refined invariants.
In each case, we relate the pyramid partitions to states in a dimer model,
in terms of which the shuffling operation (wall crossing) is most easily described.
We then explain how both refined and unrefined models are related
to the topological vertex, which consists not of pyramids
but of three-dimensional cubic ``plane partitions,'' or three-dimensional Young diagrams.

\bigskip {\noindent{\it Unrefined invariants}}

\ifig\nrERCn{The ``empty room configurations'' for the crystals that count BPS states in chambers $C_n$ and $\tilde{C}_n$.}
{\epsfxsize4.5in\epsfbox{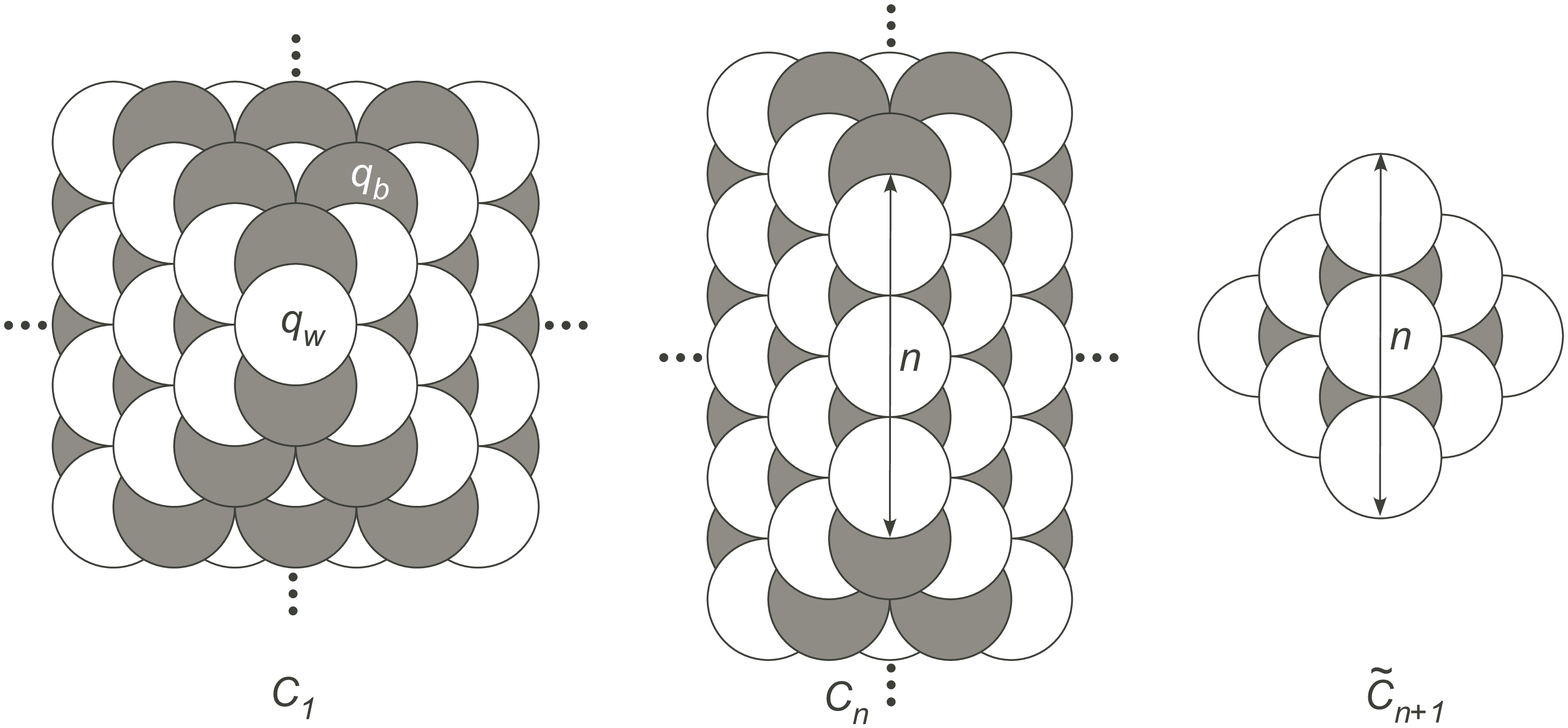}}

In the chamber $C_n$, the unrefined generating function of BPS states
is obtained by counting the melting configurations of an infinite pyramid-shaped crystal whose top row of atoms has length $n$ (sometimes also called an empty room configuration, or ERC, of length $n$) \refs{\Szendroi,\JafferisC}. As shown in \nrERCn, this crystal has two different types of atoms, corresponding to the two vertices of the Klebanov-Witten quiver. The top edge of the pyramid always consists of $n$ white atoms. The remainder of the pyramid is then constructed by placing two black atoms underneath each white atom, oriented vertically, and two white atoms underneath each black one, oriented horizontally.\foot{In nature, such a crystal structure, very similar to that of diamond, occurs in the minerals zincblende (zinc sulfide) and moissanite (silicon carbide).} In order for an atom to be removed during crystal melting, all atoms lying above it must be removed as well. The partition function is defined as a sum over all melting configurations ({\it i.e.} pyramid partitions) $\pi$,
\eqn\Cncrys{ Z(q_w,q_b;C_n) = \sum_\pi q_w^{w_w(\pi)} q_b^{w_b(\pi)}\,, }
where $w_w(\pi)$ and $w_b(\pi)$, respectively, are the numbers of white and black atoms removed. It was proven in \Young\ that this agrees with the partition function \zcn,
\eqn\CncrysDT{ Z(q_w,q_b;C_n)=Z(q,Q;C_n) = M(q)^2\prod_{j=1}^\infty(1-q^jQ)^j\prod_{k=n}^\infty(1-q^kQ^{-1})^k\,, }
provided that one makes an $n$-dependent identification as in \JafferisC,
\eqn\URvars{C_n\,:\qquad q_w = -q^n Q^{-1}\,, \qquad q_b = -q^{-(n-1)}Q \,. }

Similarly, it was argued in \JafferisC\ that to obtain the unrefined partition function in the chamber $\tilde{C}_n$ one must sum over the melting configurations of a {\it finite} crystal configuration of length $n$, also shown in  \nrERCn. Then
\eqn\Ctncrys{ Z(q,Q;\tilde{C}_{n+1}) = \prod_{j=1}^n(1-q^jQ)^j = \sum_\pi q_w^{w_w(\pi)} q_b^{w_b(\pi)} \, }
if one identifies
\eqn\URtvars{\tilde{C}_{n+1}\,:\qquad q_w = -q^nQ\,,\qquad q_b = -q^{-(n+1)}Q^{-1}\,.}

\ifig\crysdim{The relation between pyramid partitions and dimer states, illustrated for $n=2$.}
{\epsfxsize4.0in\epsfbox{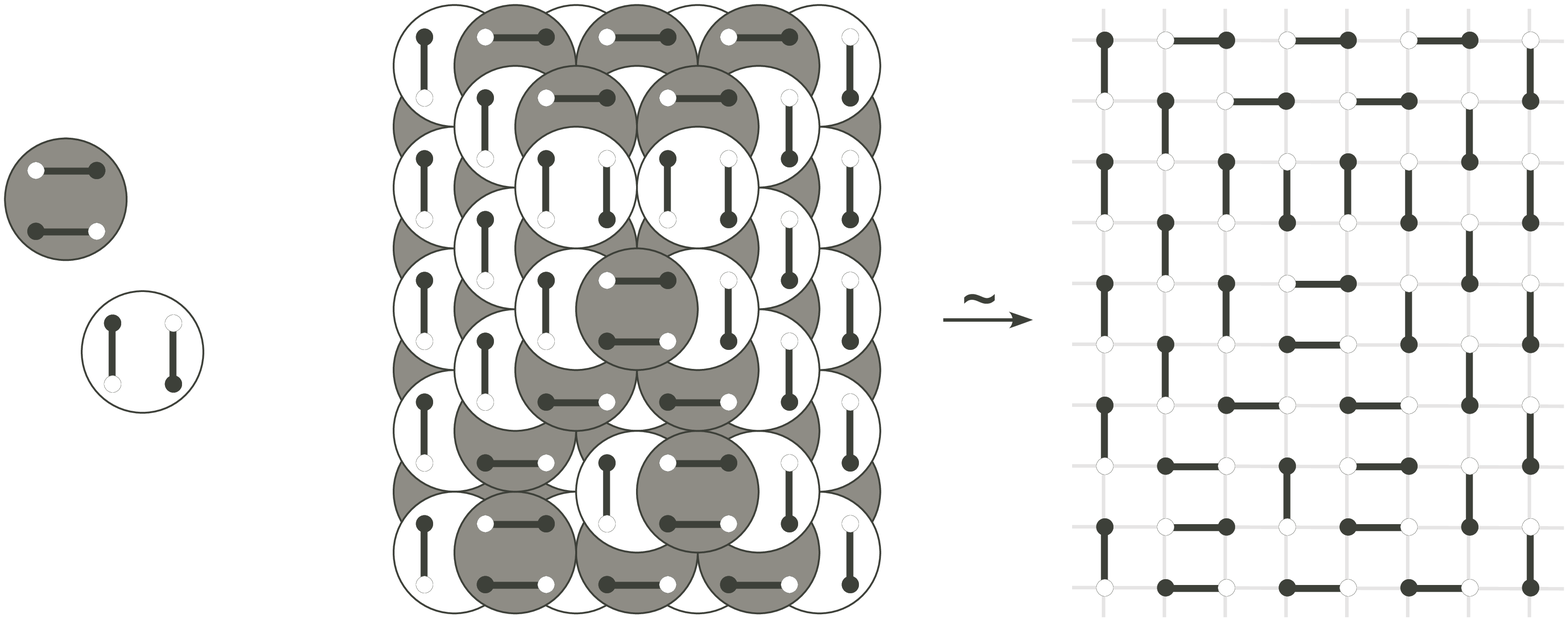}}

\ifig\evenodd{Even and odd boxes of dimers.}
{\epsfxsize2.4in\epsfbox{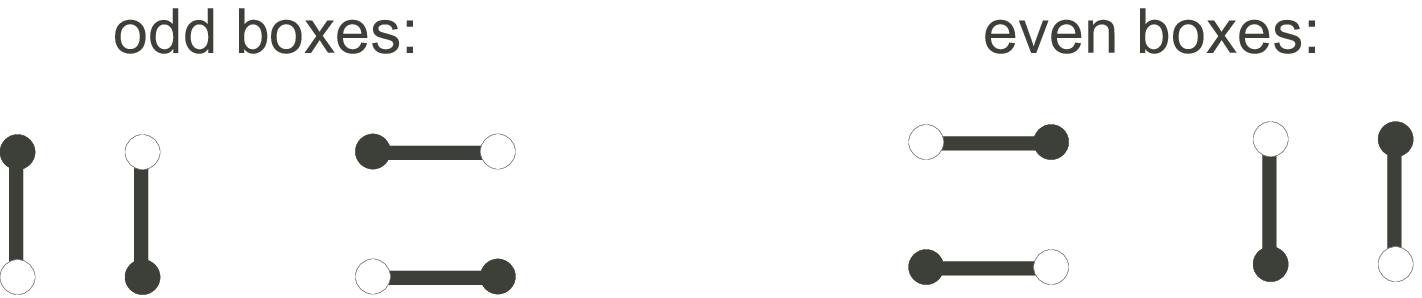}}

To proceed further, let us translate the above partition functions into the language of dimers. The partitions of a length-$n$ pyramid correspond bijectively to the states of a dimer model on a square lattice with prescribed asymptotic boundary conditions. (We will refer to these states as partitions as well.) An intuitive way to visualize the correspondence (see also \Young) is to actually draw dimers on the black and white atoms, as in  \crysdim. Then the dimer state corresponding to a given crystal automatically appears when viewing the crystal from above.

As in \Young, we have included an extra decoration on the lattices in these figures: lattice points are colored with alternating black and white dots. This canonical decoration carries no extra information, but is very useful in describing weights and wall crossing. We will also call squares in the dimer lattice {\it even} or {\it odd} depending on their vertex decorations. As shown in  \evenodd, we call two dimers lying on the edges of an even (resp. odd) square an even (resp. odd) {\it box}; an even (resp. odd) box with two horizontal (resp. vertial) dimers corresponds to a fully uncovered black (resp. white) atom in the crystal.

\ifig\weights{The weights assigned to edges of the dimer lattice of ``length $n$,'' for $n=2$. (The $n=2$ ground state has been shaded in.) All vertical edges have weight $1$ and all horizontal edges have an additional factor of $(-Q)^{-1/2}$.}
{\epsfxsize2.7in\epsfbox{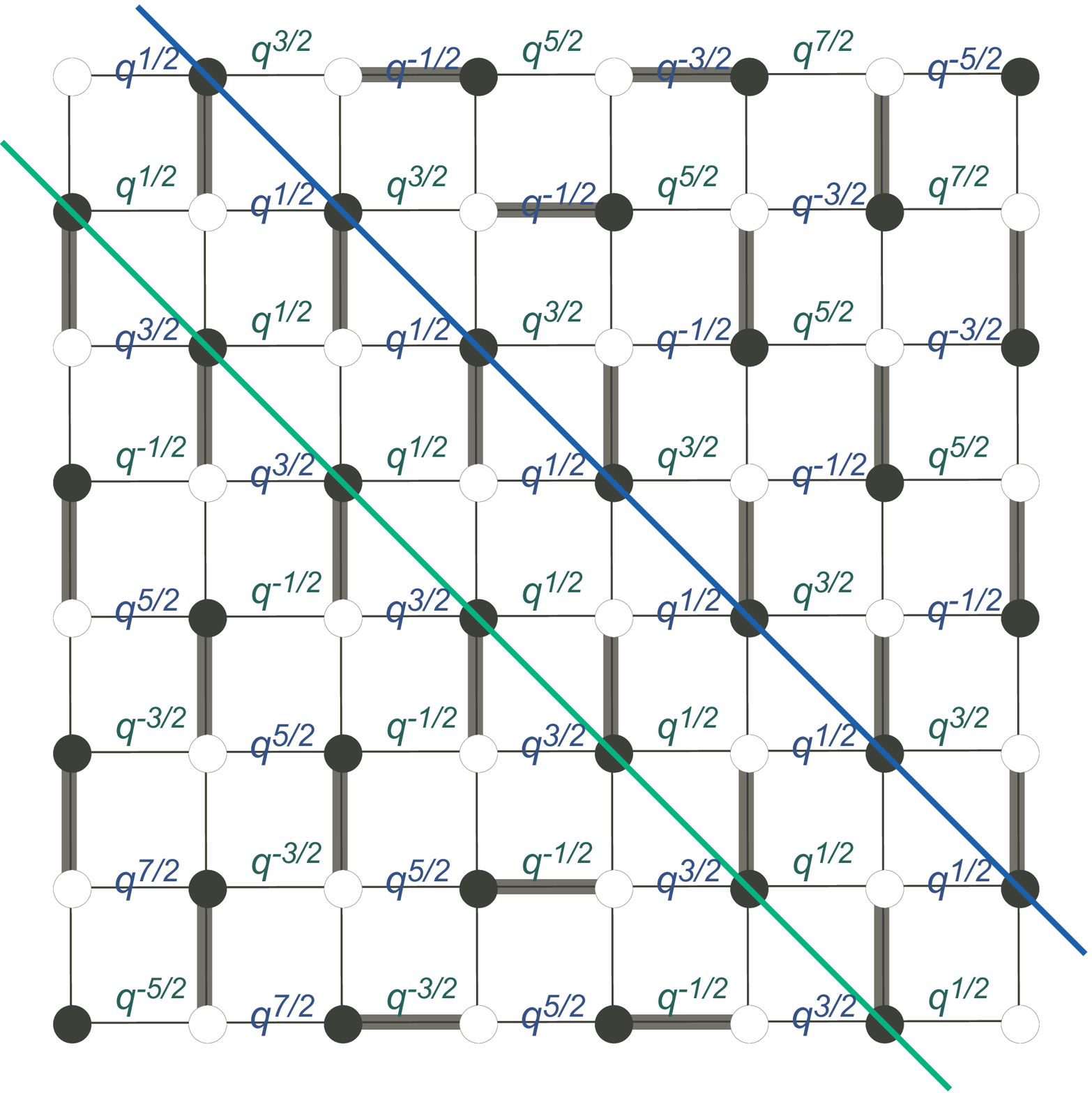}}

One can assign weights to each edge in the dimer lattice so that the total weight of a dimer partition $\pi$, defined as\foot{Technically, both the numerator and denominator in this definition must be ``regularized.'' For a given state $\pi$, one fixes a large box in the dimer lattice so that all dimers outside the box match the ground state (corresponding to an unmelted pyramid of length $n$); then one only multiplies together the weights of dimers inside this box.}
$$ w(\pi) = {{\rm product\,of\,weights\,of\,dimerized\,edges\,in}\,\pi \over {\rm product\,of\,weights\,of\,dimerized\,edges\,in\,the\,ground\,state\,of\,the\,lattice}}\,, $$
agrees with the pyramid partition weight $q_w^{w_w(\pi)} q_b^{w_b(\pi)}$.
To implement such a weighting, it is sufficient to ensure that the ratio
of horizontal to vertical edges in every odd and even square, respectively, equals $q_w$ and $q_b^{-1}$ ---
corresponding to white atoms being removed and black atoms being replaced.

Here, it is most convenient to use a weighting that is $n$-dependent. Vertical edges are always assigned weight $1$. For the horizontal edges, we draw two diagonals on the dimer lattice, which pass through the lowermost and uppermost odd blocks in the ground state dimer ({\it i.e.} the lowermost and uppermost uncovered white atoms in the unmelted pyramid). For positive integers $a$, the horizontal edges $2a-1$ units above and $2a$ units below the lower diagonal are assigned weights $q^{(2a-1)/2}(-Q)^{-1/2}$ and $q^{-(2a-1)/2}(-Q)^{-1/2}$, respectively, where $a=0$ means that an edge is touching the diagonal. Likewise the horizontal edges $2a-1$ units below and $2a$ units above the upper diagonal are assigned weights $q^{(2a-1)/2}(-Q)^{-1/2}$ and $q^{-(2a-1)/2}(-Q)^{-1/2}$, respectively. An example is shown in \weights. For a dimer model corresponding to a length-$n$ crystal, one can check that the ratios of horizontal to vertical edges in every odd block is indeed $-q^nQ^{-1}=q_w$, and in every even block the ratio is $-q^{n-1}Q^{-1}=q_b^{-1}$. Since the resulting weight function itself is $n$-dependent in terms of variables $q$ and $Q$, let us call it $w_n$ rather than $w$.

We let the weight $w_n$ be a function acting linearly on formal sums of partitions, and define $\Theta^{(n)}$ to be the formal sum of all possible partitions of a dimer lattice with asymptotic boundary conditions corresponding to the length-$n$ crystal. Then
$$ Z(q_a,q_b;C_n) = \sum_\pi q_w^{w_w(\pi)} q_b^{w_b(\pi)} = w_n(\Theta^{(n)})\,. $$

\ifig\shuffle{The directions in which dimers move under the shuffle $\tilde{S}$, and an example of shuffling a partition of length $n=2$ with odd boxes deleted.}
{\epsfxsize4.0in\epsfbox{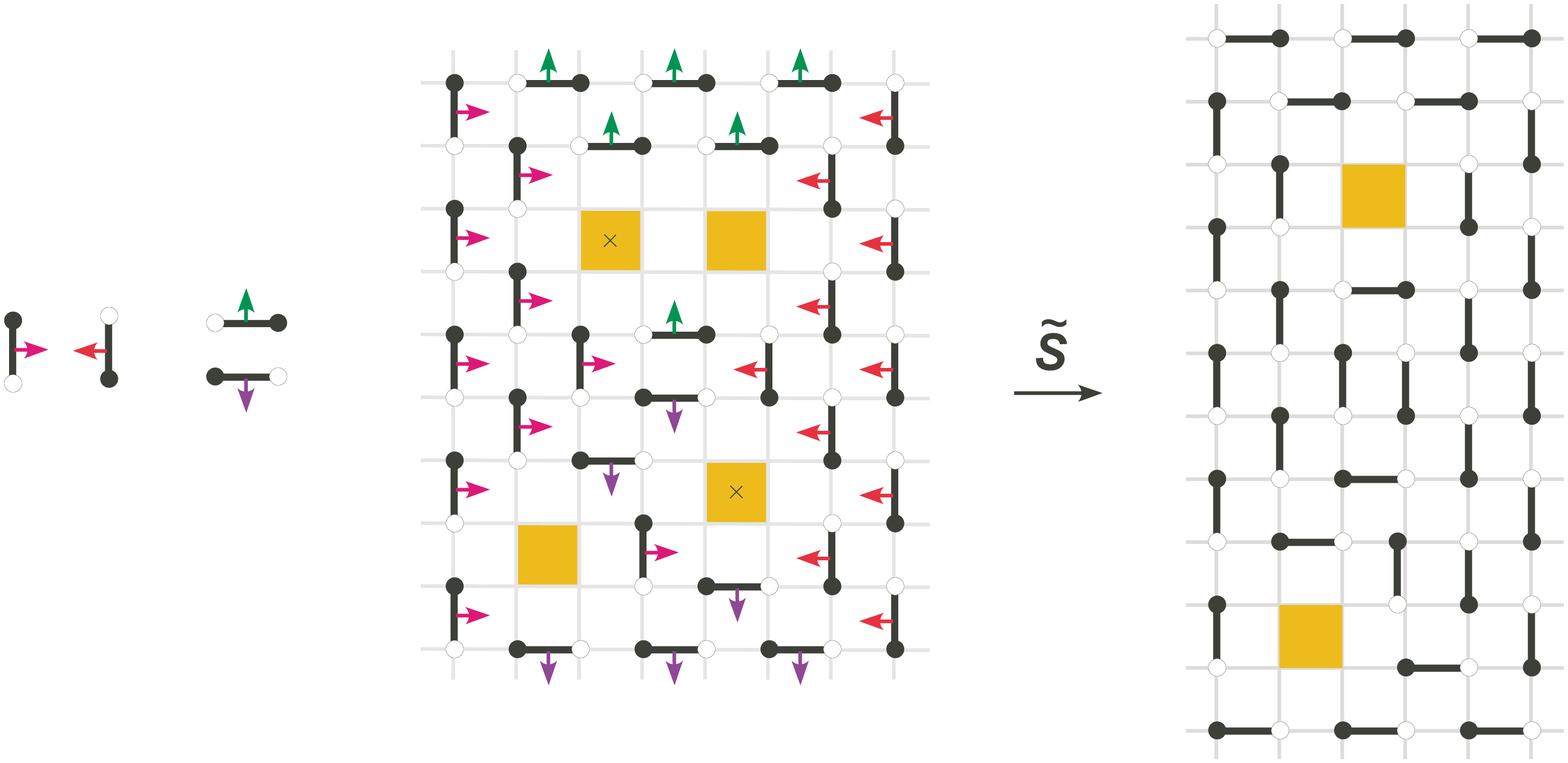}}

The operation that we claim is the combinatorial equivalent of wall crossing
is described in \Young\ as {\it dimer shuffling}.
It maps partitions of length $n$ to partitions of length $n+1$.
To define it, first consider an operation $\tilde{S}$,
which maps a dimer state $\tilde{\pi}^{(n)}$, all of whose odd blocks have been deleted,
to a dimer state $\tilde{\pi}^{(n+1)}$, all of whose even blocks are deleted.
By ``deleted'' we mean that any dimers forming odd (resp. even) blocks are removed.
The operation $\tilde{S}$ simply moves every non-deleted dimer one unit to the left,
right, up, or down, according to the rules on the left side of  \shuffle.
We show an example of such a shuffling in \shuffle\ as well;
note that dimers carry their vertex decorations with them when they move.
As a function from the set of $\{$dimer partitions with odd blocks
deleted$\}$ to the set of $\{$dimer partitions with even blocks deleted$\}$,
$\tilde{S}$ is bijective \refs{\Young,\EKLP}.
The actual dimer shuffling operation $S$ can then be defined to act on finite ``subsums'' in $\Theta^{(n)}$. It maps each formal sum%
\foot{We could also define shuffling, as in \Young, to act on individual $\pi$'s, but this is unnecessary.}
 of $2^m$ dimer states with a fixed set of $m$ odd blocks (for any $m$)
to the finite formal sum of all dimer states with a fixed set of even blocks in the obvious way:
by deleting odd blocks, applying $\tilde{S}$, and filling in the missing even blocks in all possible combinations.
Letting $S$ act linearly on all such formal sub-sums of $\Theta^{(n)}$, it must, because $\tilde{S}$ is bijective, send $\Theta^{(n)}$ precisely to $\Theta^{(n+1)}$.

What happens to weights under dimer shuffling?
We defined our weight function above so that the dimers in a partition $\tilde{\pi}$
of a length-$n$ model with odd boxes deleted do not change weight at all under the action of $\tilde{S}$.
In other words,
$w_n(\tilde{\pi})=w_{n+1}(\tilde{S}(\tilde{\pi}))$.
The only change in weights of a genuine dimer state $\pi$
under the action of $S$ arises from the deletion of odd blocks
and the subsequent creation of new even blocks after shuffling.
An important lemma in \Young\ (which we will refine later in this section)
is that the difference between the number of deleted odd blocks in $\tilde{\pi}$
and the missing even blocks in $\tilde{S}(\tilde{\pi})$ is always exactly $n$.
Then a quick exercise shows that for a fixed $\tilde{\pi}$ with $m$ deleted odd blocks,
$$
\eqalign{
w_n({\rm sum\,of\,}\pi{\rm \,s.t.}\,\pi\,{\rm agrees\,with}\,\tilde{\pi})
& = (1-q^nQ^{-1})^m \cdot w_n(\tilde{\pi})\,, \cr
w_{n+1}({\rm sum\,of\,}\pi{\rm \,s.t.}\,\pi\,{\rm agrees\,with}\,\tilde{S}(\tilde{\pi}))
& = (1-q^nQ^{-1})^{m-n}\cdot w_{n+1}(\tilde{S}(\tilde{\pi}))\,. }
$$
(By ``agrees with,'' we mean aside from deleted blocks.)
The ratio of these quantities is independent of $m$, immediately proving that
\eqn\URdWCF{ w_n(\Theta^{(n)}) = (1-q^nQ^{-1})^n w_{n+1}(\Theta^{(n+1)})\,. }
This is precisely the wall crossing formula between chambers $C_n$ for the conifold.

Formula \URdWCF\ suggests (correctly) that we can  write the crystal
or dimer partition function for a model of length $n$ as
$$  w_n(\Theta^{(n)}) = \prod_{j=n}^\infty (1-q^jQ^{-1})^j \,\cdot w_\infty(\Theta^{(\infty)})\,. $$
Of course, the quantity $w_\infty(\Theta^{(\infty)})$ must be
the Donaldson-Thomas partition function of the conifold,
and this relation holds because pyramid partitions
of length $n\rightarrow\infty$ effectively reduce to
the topological vertex formalism of \refs{\AKMV,\ORV} (see also \refs{\INOV,\MNOP}).

\ifig\brick{The brick-like lattices around the upper and lower vertices as $n\rightarrow\infty$.
The ground state state of the dimer is shaded in.
As before, each horizontal edge also carries a weight of $(-Q)^{-1/2}$}
{\epsfxsize5.0in\epsfbox{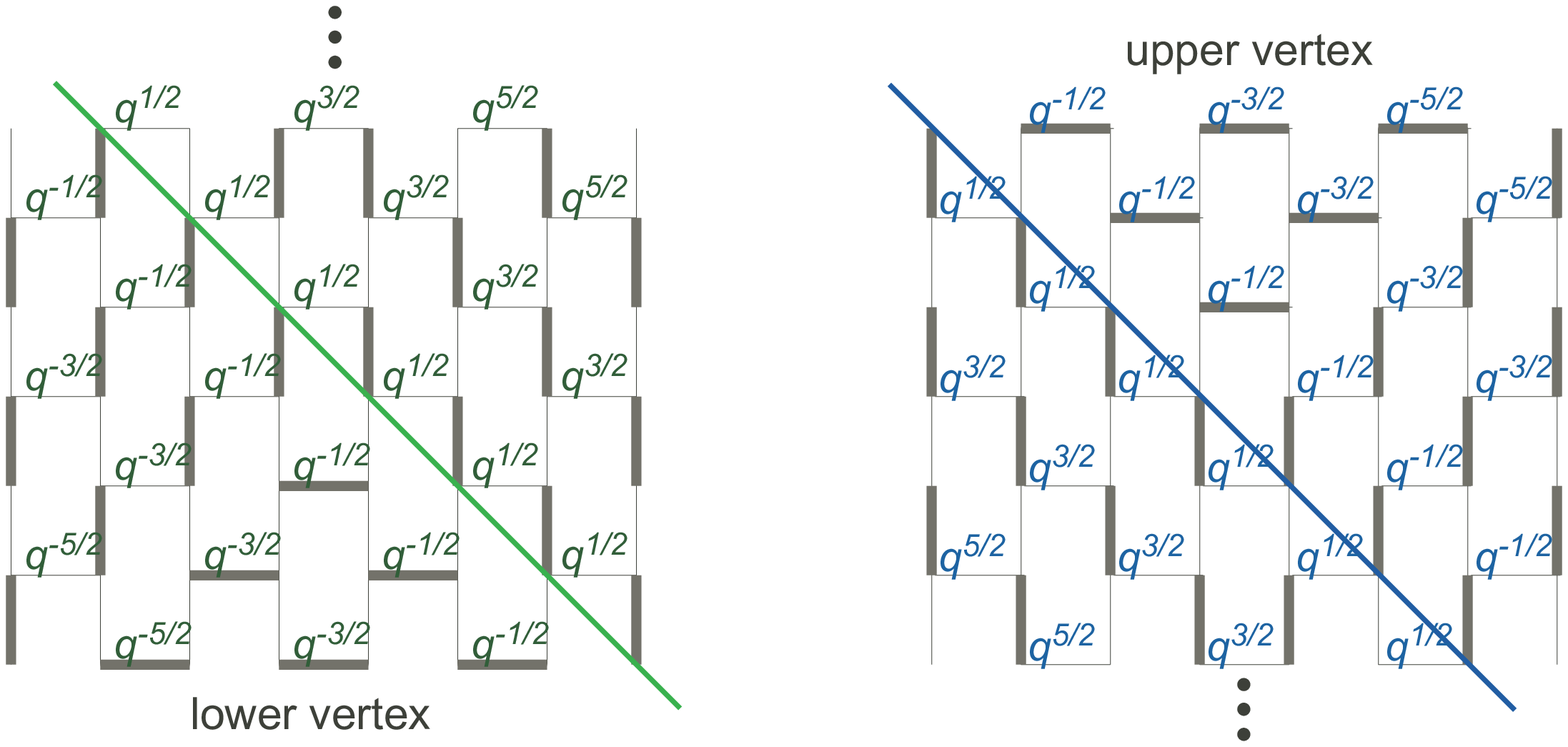}}

To understand this relation, consider the $n$-dependent weighting system of  \weights.
In the limit $n\rightarrow\infty$, the weights of half the edges around the lower
vertex (of the pyramid, or of the dimer model) acquire infinitely large, positive powers of $q$ and cease to contribute
to the partition function. Likewise for half the edges around the upper vertex.
Therefore, the only dimer partitions around these vertices
that can contribute to the length-infinity partition function involve dimers
on edges of the brick-like lattices of  \brick. These brick-like lattices, however,
are equivalent to hexagonal dimer lattices, which correspond to the three-dimensional
cubic partitions that arise in the topological vertex.

\ifig\topvx{The map between the length-infinity dimer model and a pair of topological vertices.
(The extra $q_1$ and $q_2$ notations are for the refined case further below.)}
{\epsfxsize4.0in\epsfbox{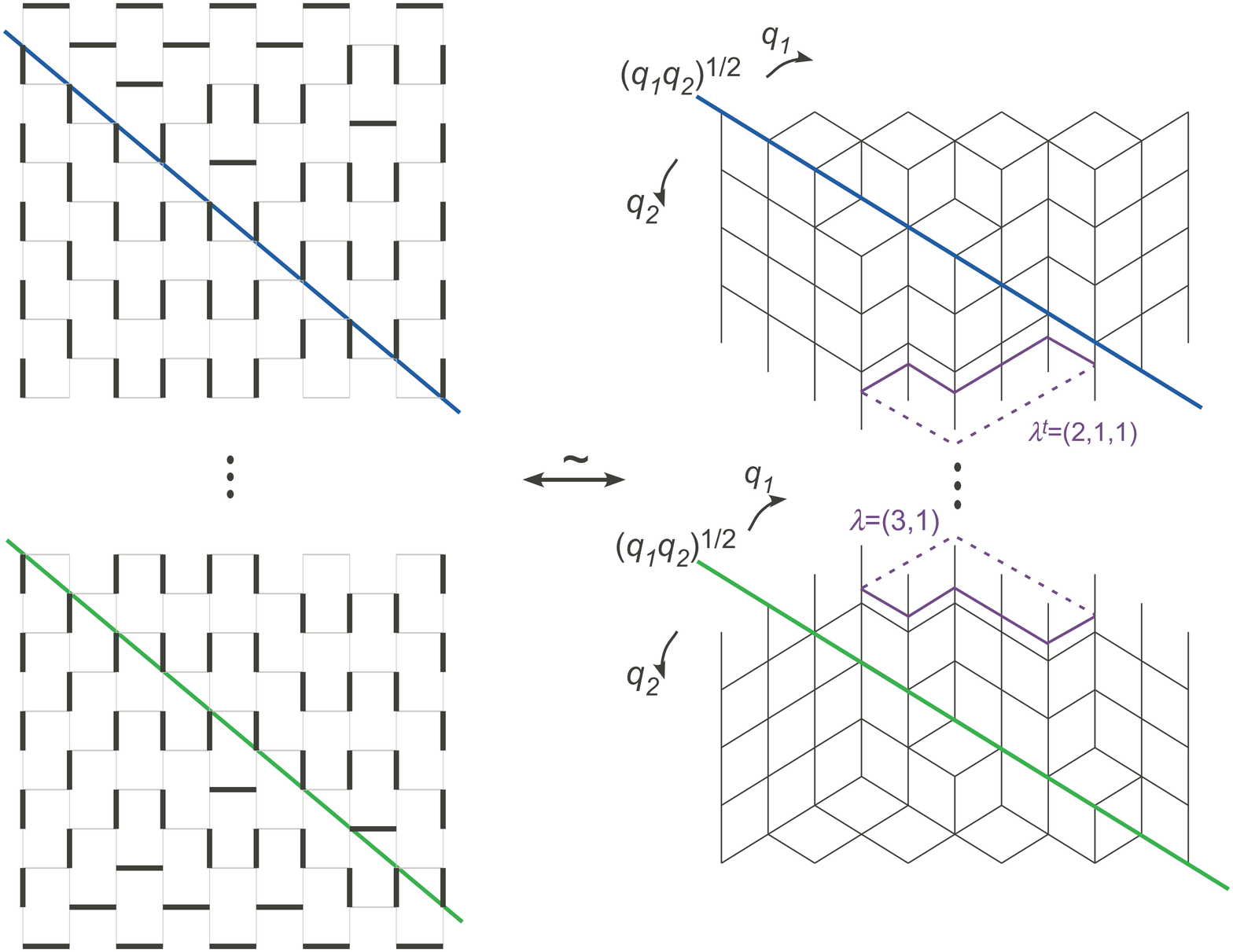}}

As argued more carefully in \Young, any (nontrivial) configuration of the length-infinity
dimer model can be constructed via a series of moves that amount
to 1) cutting out a Young diagram $\lambda$ simultaneously from the upper
and lower vertices, 2) stacking up individual boxes to form a cubic
partition $\pi_\lambda^-$ around the lower vertex, and 3) stacking up
boxes to form a partition $\pi_\lambda^+$ around the upper vertex.
An example of such a dimer configuration and its corresponding topological vertex
partitions is shown in  \topvx. By observing how dimers shift in these three
steps and using our $n\rightarrow\infty$ weighting, it is not too hard to see
that the contributions to the partition function are
$(-Q)^{|\lambda|}  q^{{1\over2}||\lambda||^2}q^{{1\over2}||\lambda^t||^2}$ from
step (1), $q^{|\pi_\lambda^-|}$ from step (2), and $q^{|\pi_{\lambda^t}^+|}$ from
step (3).\foot{We use conventional notation for Young diagrams and
three-dimensional cubic partitions; $\lambda^t$ is the transpose of the diagram $\lambda$,
the rows of $\lambda$ have lengths $\lambda_i$, $|\lambda| = \sum \lambda_i$
is the number of boxes in $\lambda$, $||\lambda||^2 = \sum \lambda_i^2$,
and $|\pi|$ is the number of boxes in a three-dimensional partition.} Therefore, the total partition function is
$$ w_\infty(\Theta^{(\infty)}) = \sum_\lambda \sum_{\pi_{\lambda^t}^+,\,\pi_\lambda^-} (-Q)^{|\lambda|}  q^{{1\over2}||\lambda||^2+{1\over2}||\lambda^t||^2} q^{|\pi_{\lambda^t}^+|+|\pi_\lambda^-|}\,, $$
which is {\it precisely} the topological vertex expression for the (unreduced)
partition function of the conifold \refs{\AKMV,\ORV}.
In terms of Schur functions, the generating function for three-dimensional
cubic partitions with a single nontrivial asymptotic boundary condition $\lambda$
is  $\sum_{\pi_\lambda}q^{|\pi_\lambda|} = M(q)\,q^{-{1\over2}||\lambda||^2}s_{\lambda^t}(q^{-\rho}) = M(q)\,q^{-{1\over2}||\lambda||^2}s_{\lambda^t}(q^{1/2},q^{3/2},q^{5/2},...)$. Thus, as expected,
\eqn\wtconif{\eqalign{
w_\infty(\Theta^{(\infty)}) & = Z(q,Q;C_\infty) \cr
& = M(q)^2\sum_\lambda (-Q)^\lambda s_\lambda(q^{-\rho})s_{\lambda^t}(q^{-\rho}) \cr
& = M(q)^2 \prod_{j,k=1}^\infty (1-q^{j-1/2}q^{k-1/2}Q) \cr
& = M(q)^2 \prod_{j=1}^\infty (1-q^jQ)^j\,. }}
%

%%%%%%%%%%%%%%%%%%%%%%%%%%%%%%%%%%%%%%%%%%%%%%%%%%%%%%%%%%%%%%%%%%%%%%%%%%%%%%%%

\bigskip\noindent {\it Refined invariants}

We now come to the crystal melting models for refined invariants. For the conifold, we describe models to compute the refined partition functions in all chambers $C_n$ and $\tilde{C}_n$. We will first present the formulae in terms of melting crystals, and then prove them while discussing their relation to dimer shuffling, refined wall crossing, and the the refined topological vertex.

\ifig\RERC{Weights of atoms for the refined partition function in chamber $C_1$.}
{\epsfxsize2.0in\epsfbox{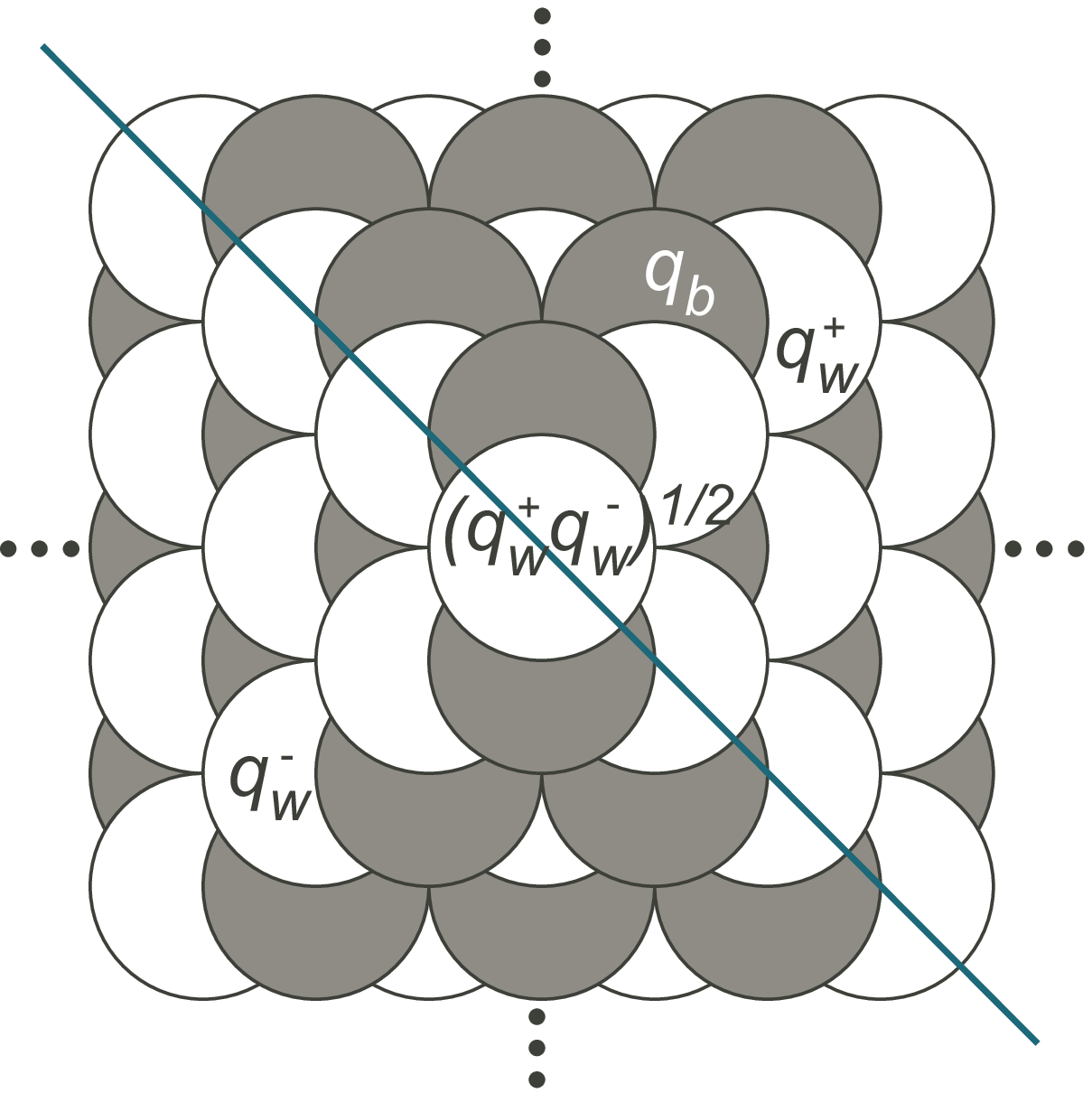}}

At the level of crystal models, one must draw a series of diagonals on the pyramid partition,
and interpolate weights between the variable $q_1$ on one side of the diagonals and $q_2$ on the other.
To be more specific, consider the pyramid of length $n=1$, corresponding to the Szendr\"oi chamber $C_1$. On this crystal model, we draw a single diagonal
as shown in \RERC; we assign white atoms above the diagonal
a weight $q_w^+$, white atoms below the diagonal a weight $q_w^-$,
and white atoms on the diagonal itself a weight $(q_w^+q_w^-)^{1/2}$.
All black atoms are assigned weight $q_b$.
Letting $w_w^+(\pi)$, $w_w^-(\pi)$, and $w_w^0(\pi)$ be the numbers of white atoms
above, below, and on the diagonal, respectively, in the partition $\pi$,
and identifying $q_w^+ = -q_1Q^{-1}$, $q_w^-=-q_2Q^{-1}$, and $q_b=-Q$, we find
\eqn\zwbref{\eqalign{
Z(q_w^+,q_w^-,q^b;C_1)
& = \sum_\pi (q_w^+)^{w_w^+(\pi)}(q_w^-)^{w_w^-(\pi)}(q_w^+q_w^-)^{{1\over2}w_w^0(\pi)}q_b^{w_b(\pi)} \cr
& = Z^{\rf} (q_1,q_2,Q;C_1) \cr
& = M(q_1,q_2)^2\prod_{i,j=1}^\infty (1-q_1^{i-{1\over2}}q_2^{j-{1\over2}}Q)(1-q_1^{i-{1\over2}}q_2^{j-{1\over2}}Q^{-1})\,. }}

\ifig\RERCn{Refined weights of atoms for chambers $C_n$ and $\tilde{C}_{n+1}$, with $n=3$.}
{\epsfxsize5.0in\epsfbox{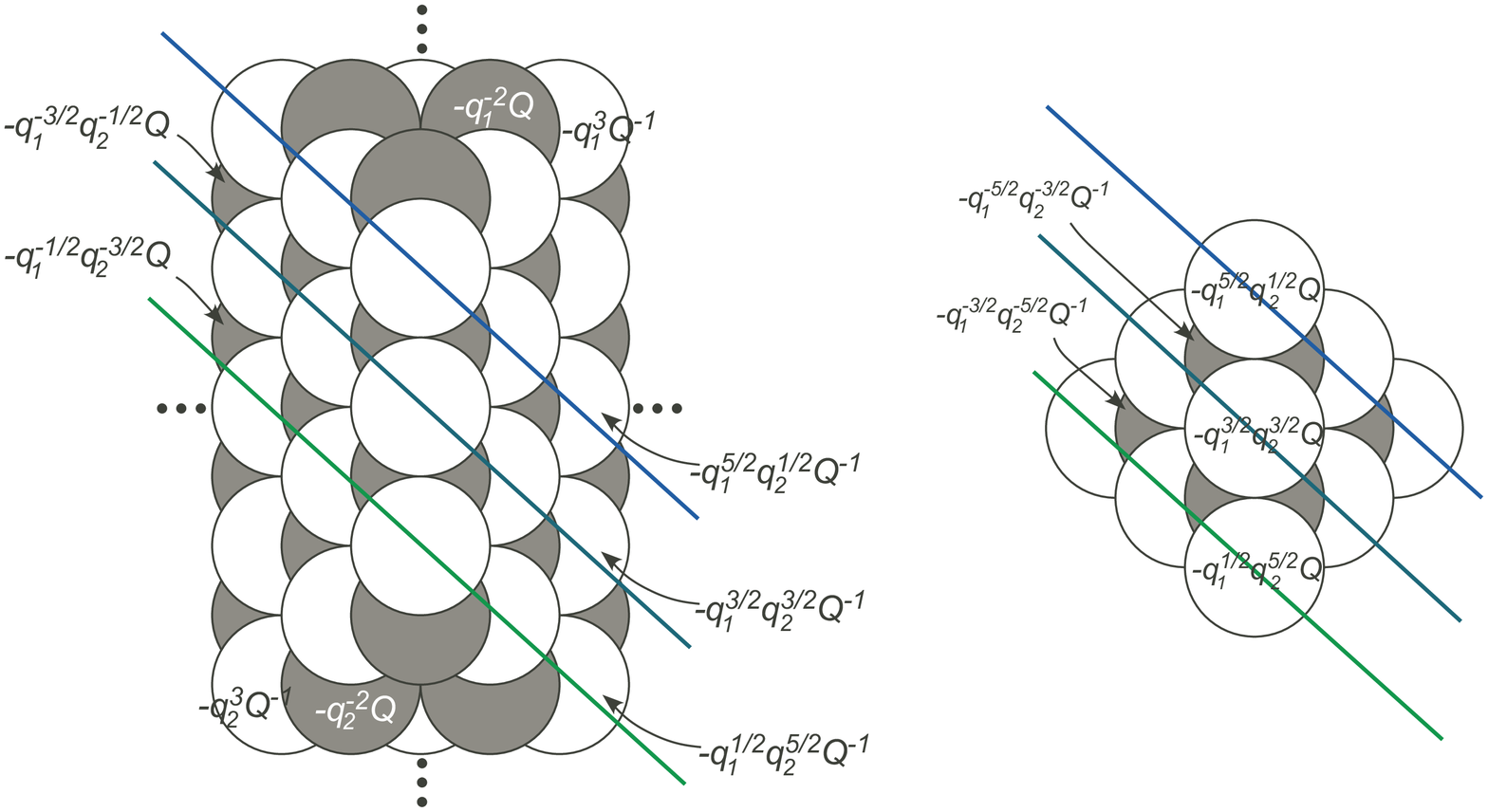}}

To generalize to the length-$n$ pyramid, we draw $n$ diagonals, as in the left half of \RERCn.
It is more natural to work directly in terms of the variables $q_1$, $q_2$, and $Q$.
We assign weights $-q_1^nQ^{-1}$ (resp. $-q_1^{-(n-1)}Q$) to the white (resp. black)
atoms above all the diagonals and weights $-q_2^nQ^{-1}$ (resp. $-q_2^{-(n-1)}Q$)
to the white (resp. black) atoms below all the diagonals.
The diagonals themselves intersect white atoms; we assign the same weight
to all the white atoms on a single diagonal, interpolating between
$-q_1^{n-{1\over2}}q_2^{1\over2}Q^{-1}$ on the uppermost diagonal and $-q_1^{{1\over2}}q_2^{n-{1\over2}}Q^{-1}$
on the lowermost (multiplying by $q_1^1q_2^{-1}$ in each intermediate step).
Similarly, black atoms lie between diagonals, and we assign them
weights ranging from $-q_1^{-n+{3\over2}}q_2^{-{1\over2}}Q$
directly below the upper diagonal to $-q_1^{-1\over2}q_2^{-n+{3\over2}}Q$
directly above the lower diagonal.
Multiplying together the weights of all atoms removed in a given partition $\pi$
and summing these quantities over partitions, we obtain the expected
\eqn\RcrysCn{
Z^{\rf} (q_1,q_2,Q;C_n) = M(q_1,q_2)^2 \prod_{i,j=1}^\infty (1-q_1^{i-{1\over2}}q_2^{j-{1\over2}}Q)
\prod_{i\geq 1,\,j\geq 1\atop i+j>n} (1-q_1^{i-{1\over2}}q_2^{j-{1\over2}}Q^{-1})\,. }

For chambers $\tilde{C}_{n+1}$,
the finite pyramid of length $n$ can also be split by $n$ diagonals,
as shown in the right half of  \RERCn. If one assigns weights such
that (1) when $q_1\rightarrow q$ and $q_2\rightarrow q$ white atoms have weight $-q^nQ$
and black atoms have weight $-q^{-(n+1)}Q^{-1}$; (2) when moving up one step,
either on or inbetween diagonals, the absolute value of the power of $q_2$ (resp. $q_1$) decreases (resp. increases) by $1$; and (3) the assignment is symmetric about the middle diagonal(s) of the crystal, the resulting partition function is precisely
$$
Z^{\rf} (q_1,q_2,Q;\tilde{C}_{n+1})
= M(q_1,q_2)^2 \prod_{i\geq 1,\,j\geq 1\atop i+j\leq n+1} (1-q_1^{i-{1\over2}}q_2^{j-{1\over2}}Q )\,.
$$

\ifig\weightsR{Refined weighting of the length-$n$ dimer, for $n=2$. (The color coding here differs from that in \weights, to emphasize the difference between $q_1$ and $q_2$.)}
{\epsfxsize2.7in\epsfbox{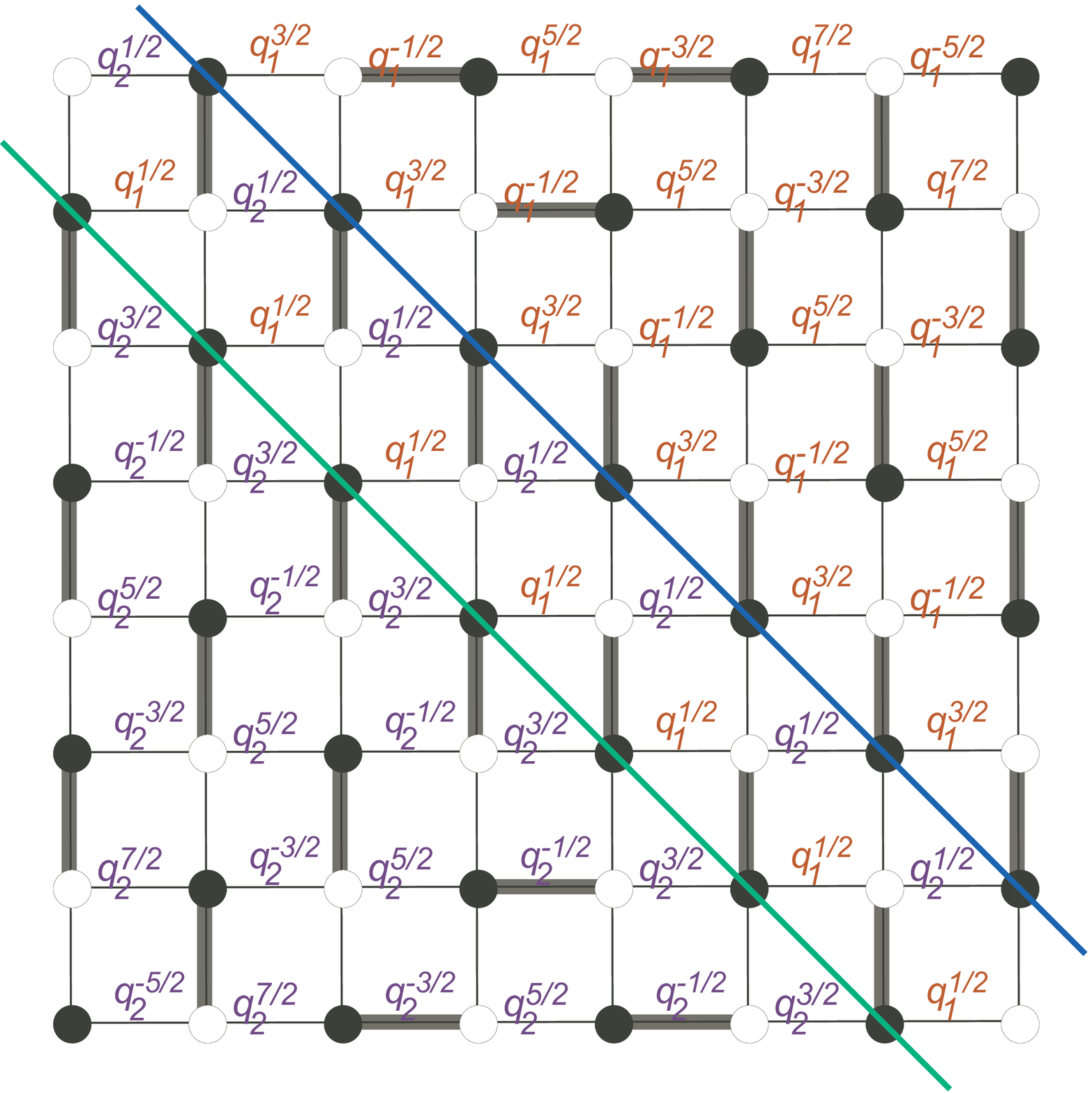}}

For the remainder of the section we return to the infinite pyramid of length $n$,
generalizing the previous unrefined discussion to refine the connection between shuffling, wall crossing, and the refined topological vertex (and to prove formula \RcrysCn).
We first observe that in order to equate refined pyramid partitions and their weights with states (configurations) of a dimer lattice, we can use almost the same $n$-dependent weighting described in  \weights. Now, for positive integers $a$, the horizontal edges $2a-1$ units above and $2a$ units below the lower diagonal are assigned weights $q_1^{(2a-1)/2}(-Q)^{-1/2}$ and $q_2^{-(2a-1)/2}(-Q)^{-1/2}$, respectively. Likewise the horizontal edges $2a-1$ units below and $2a$ units above the upper diagonal are assigned weights $q_2^{(2a-1)/2}(-Q)^{-1/2}$ and $q_1^{-(2a-1)/2}(-Q)^{-1/2}$. See the example in  \weightsR.

As in the unrefined case, the weights of dimers which are not part of deleted odd or even blocks do not change during dimer shuffling $\tilde{S}$, due to our $n$-dependent weighting. In order to understand the behavior of the deleted blocks, we observe that the shuffling $\tilde{S}$ removes {\it exactly} one (deleted) odd block from each of the $n$ diagonals of a dimer configuration of length $n$. Moreover, the remaining (deleted) odd blocks are mapped to deleted even blocks with exactly the same weights --- if (for instance) they were above all the diagonals, then they remain above all the diagonals. These statements can be proved with careful counting arguments, considering the number of dimers on and around each diagonal in an arbitrary configuration before and after shuffling. The result is that when the actual shuffling $S$ maps a formal sum of states $\pi$ agreeing with a fixed odd-deleted state $\tilde{\pi}$ on all but their odd blocks to a formal sum of states agreeing on all but their even blocks, the weight of this formal sum changes by exactly $\prod_{i+j=n+1} (1-q_1^{i-{1\over2}}q_2^{j-{1\over 2}}Q^{-1})$; therefore,
$$  w_n(\Theta^{(n)}) = \prod_{i\geq 1,\, j\geq 1\atop i+j=n+1} (1-q_1^{i-{1\over2}}q_2^{j-{1\over 2}}Q^{-1})\,\cdot w_{n+1}(\Theta^{(n+1)})\,. $$
This, of course, is the refined wall crossing formula for chambers $C_n$.

\ifig\brickR{Neighborhoods of the refined upper and lower vertices as $n\rightarrow\infty$.}
{\epsfxsize5.0in\epsfbox{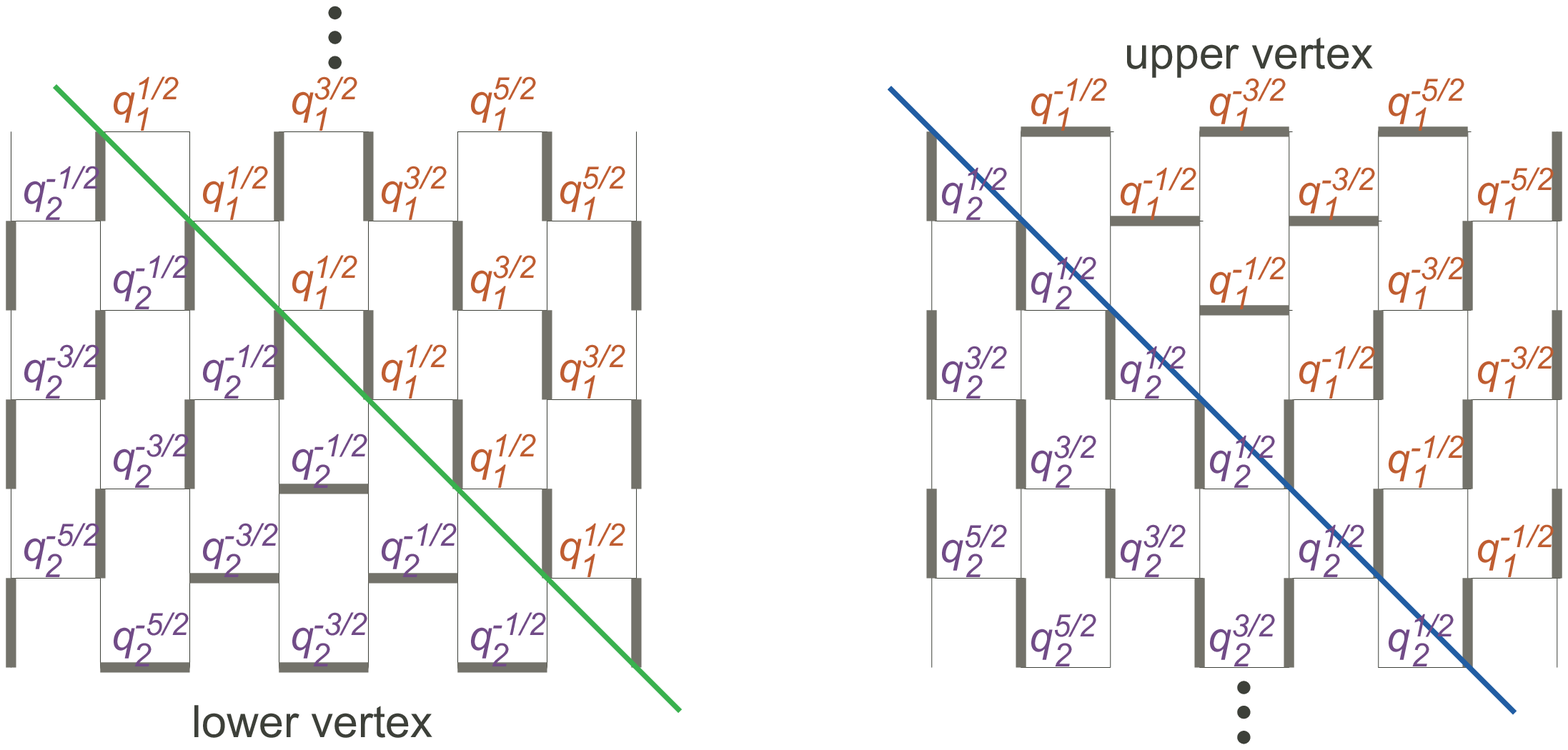}}

The crystal-melting or dimer partition function of length $n$ can now be written as
$$ w_n(\Theta^{(n)}) = \prod_{i,j=1}^\infty (1-q_1^{i-{1\over2}}q_2^{j-{1\over 2}}Q^{-1})\,\cdot w_{\infty}(\Theta^{(\infty)})\,. $$
The last term, $w_{\infty}(\Theta^{(\infty)})$, is obtained from a slightly modified version of the refined topological vertex of \IKV. To see this, observe that as $n\rightarrow\infty$ the neighborhoods of the upper and lower vertices of the dimer lattice still reduce to effective brick-like lattices, now shown in  \brickR. In terms of three-dimensional cubic partitions, states of the length-infinity dimer are again created by 1) cutting out a Young diagram $\lambda$ simultaneously from the upper and lower vertices, 2) stacking up individual boxes to form a cubic partition $\pi_\lambda^-$ around the lower vertex, and 3) stacking up boxes to form a partition $\pi_\lambda^+$ around the upper vertex. The creation of the Young diagram $\lambda$ comes with a fairly simple weight $(-Q)^{|\lambda|}q_1^{{1\over2}||\lambda||^2}q_2^{{1\over2}||\lambda^t||^2}$.  However, both the upper and lower ``room corners'' are now split along a diagonal, as shown in  \topvx. In the case of the lower corner, boxes stacked below the diagonal come with weight $q_2$, those above the diagonal with weight $q_1$, and those that the diagonal intersects have weight $(q_1q_2)^{1\over2}$. The situation is reversed for the upper vertex. The generating function for such three-dimensional cubic partitions with one asymptotic boundary condition $\lambda$ is (for example, at the lower vertex)
\eqn\Rtvx{   \sum_{\pi_\lambda} q_1^{|\pi_\lambda|^{(q_1)}}q_2^{|\pi_\lambda|^{(q_2)}}(q_1q_2)^{{1\over2}|\pi_\lambda|^{(0)}} = M(q_1,q_2)\,q_2^{-{1\over2}||\lambda^t||^2}s_\lambda(q_2^{-\rho})\,, }
with $M(q_1,q_2)=\prod_{i,j=1}^\infty(1-q_1^{i-{1\over2}}q_2^{j-{1\over2}})^{-1}$.
Therefore, the length-infinity pyramid partition function is
$$
\eqalign{
w_\infty(\Theta^{(\infty)}) &=
 \sum_\lambda \sum_{\pi_{\lambda^t}^+\,,\pi_\lambda^-} (-Q)^{|\lambda|}\,q_1^{{1\over2}||\lambda||^2} q_2^{{1\over2}||\lambda^t||^2} \cr
 & \qquad\qquad \times q_1^{|\pi_\lambda^-|^{(q_1)}}q_2^{|\pi_\lambda^-|^{(q_2)}}(q_1q_2)^{{1\over2}|\pi_\lambda^-|^{(0)}}\,
q_2^{|\pi_{\lambda^t}^+|^{(q_2)}}q_1^{|\pi_{\lambda^t}^+|^{(q_1)}}(q_1q_2)^{{1\over2}|\pi_{\lambda^t}^+|^{(0)}} 
 \cr
& = M(q_1,q_2)^2\sum_\lambda (-Q)^{|\lambda|} s_\lambda(q_2^{-\rho})s_{\lambda^t}(q_1^{-\rho}) \cr
& = M(q_1,q_2)^2\prod_{i,j=1}^\infty (1-q_1^{i-{1\over2}}q_2^{j-{1\over2}}Q)\,. }
$$

Note that expression \Rtvx\ differs only slightly from the refined topological vertex
used in \IKV\ (with the boundary condition $\lambda$ placed along an ``unpreferred'' direction). The difference comes from our symmetric choice of normalization, as discussed in section 2.
In \IKV\ the diagonal is assigned to $q_2$ rather than $(q_1q_2)^{1\over2}$,
resulting in the fact that the refined MacMahon function appearing in the analogue of \Rtvx\
is not $M(q_1,q_2)=\prod_{i,j=1}^\infty(1-q_1^{i-{1\over2}}q_2^{j-{1\over2}})^{-1}$,
but rather $\prod (1-q_1^{i-1}q_2^j)^{-1}$.
The refined A-model (Gromov-Witten) partition functions calculated with
the refined vertex are always normalized by the prefactor $M(q_1,q_2)^\chi$,
so in many previous calculations this has made no difference.

%%%%%%%%%%%%%%%%%%%%%%%%%%%%%%%%%%%%%%%%%%%%%%%%%%%%%%%%%%%%%%%%%%%%%%%

\vskip 30pt

\centerline{\bf Acknowledgments}

\noindent
We thank A.~Gorsky, E.~Gorsky, D.~Jafferis, G.~Moore, A.~Neitzke, H.~Ooguri, Y.~Soibelman, and M.~Yamazaki
for useful discussions and comments.
We are grateful to the KITP, Santa Barbara for warm
hospitality during the program ``Fundamental Aspects of Superstring Theory,''
where part of this work was carried out.
TD acknowledges support from a National Defense Science and Engineering Graduate Fellowship.
Research of SG is supported in part by the Alfred P. Sloan Foundation,
by DARPA under Grant No. HR0011-09-1-0015,
and by the National Science Foundation under
Grant No. PHY05-51164 and Grant No. PHY07-57647.
Opinions and conclusions expressed here are those of the authors
and do not necessarily reflect the views of funding agencies.

\listrefs
\end